\documentclass[published]{JHEP3} 

\JHEP{00(2007)000}

\JHEPspecialurl{http://jhep.sissa.it/JOURNAL/JHEP3.tar.gz}

\usepackage{epsfig,multicol,bbm}
\usepackage{dsfont}

\newcommand\fverb{\setbox\fverbbox=\hbox\bgroup\verb}
\newcommand\fverbdo{\egroup\medskip\noindent%
            \fbox{\unhbox\fverbbox}\ }
\newcommand\fverbit{\egroup\item[\fbox{\unhbox\fverbbox}]}
\newbox\fverbbox

\newcommand{\vx}{\ensuremath{\vec{x}}}
\newcommand{\vk}{\ensuremath{\vec{k}}}

\newcommand{\be}{\begin{equation}}
\newcommand{\ee}{\end{equation}}
\newcommand{\bea}{\begin{eqnarray}}
\newcommand{\eea}{\end{eqnarray}}

\title{Sterile neutrino production via active-sterile oscillations:  the quantum Zeno effect.}

\author{D. Boyanovsky\\Department of Physics and
Astronomy, University of Pittsburgh,\\ Pittsburgh, Pennsylvania
15260, USA \\E-mail \email{boyan@pitt.edu}}

\author{C. M. Ho\\ Department of Physics and Astronomy, University of
Pittsburgh,\\ Pittsburgh, Pennsylvania 15260, USA\\E-mail
\email{cmho@phyast.pitt.edu}}

\received{May 13, 2007}       
\accepted{\today}       

\abstract{We study several aspects of the kinetic approach to
sterile neutrino production via active-sterile mixing. We obtain the
neutrino propagator in the medium including self-energy corrections
up to $\mathcal{O}(G^2_F)$, from which we extract the dispersion
relations and damping rates of the propagating modes. The dispersion
relations are the usual ones in terms of the index of refraction in
the medium, and the damping rates are  $\Gamma_1(k) = \Gamma_{aa}(k)
\cos^2\theta_m(k); \Gamma_2(k) = \Gamma_{aa}(k) \sin^2\theta_m(k)$
where $\Gamma_{aa}(k)\propto G^2_F\,k T^4$ is the active neutrino
scattering rate and $\theta_m(k)$ is the mixing angle in the medium.
We provide a generalization of the transition probability
  in  the \emph{medium from expectation values in the
density matrix}: $  P_{a\rightarrow s}(t) =
\frac{\sin^22\theta_m}{4}\left[e^{-\Gamma_1t} + e^{-\Gamma_2
t}-2e^{-\frac{1}{2}(\Gamma_1+\Gamma_2)t}~\cos\big(\Delta E
t\big)\right]  $ and study the conditions for its quantum Zeno
suppression directly in real time.  We find the general conditions
for quantum Zeno suppression, which for $m_s\sim \textrm{keV}$
sterile neutrinos with $\sin2\theta \lesssim 10^{-3}$ \emph{may only
be} fulfilled   near an MSW resonance. We discuss the implications
for sterile neutrino production and argue that in the early Universe
the wide separation of relaxation scales far away from MSW
resonances suggests the breakdown of the current kinetic approach.}


\begin{document}

\section{Introduction}\label{sec:intro}

Sterile neutrinos, namely weak interaction singlets,  are ubiquitous
in extensions of the standard
model\cite{book1,book2,book3,raffeltbook} and are emerging as
plausible cold or warm dark matter
candidates\cite{dodelson,asaka,shi,kev1,hansen,kev2,kev3,kusenko,kou,dolgovrev,pastor,hannestad,bier},
    as potentially important ingredients in stellar
collapse and supernovae\cite{raffeltSN,fuller} and in primordial
nucleosynthesis\cite{fuller2,fuller3}. Sterile neutrinos with masses
in the range $\sim \mathrm{keV}$ may also provide an explanation of
pulsar ``kicks'' via asymmetric neutrino
emission\cite{segre,fullkus}.

The MiniBooNE collaboration\cite{miniboone} has recently reported
results in contradiction with   those from LSND\cite{lsnd1,lsnd2}
that suggested a   sterile neutrino with $\Delta m^2 \sim
1~\textrm{eV}^2$ scale. Although the MiniBooNE results  hint at an
excess of events below $475~\mathrm{MeV}$ the analysis distinctly
excludes two neutrino appearance-only from $\nu_\mu \rightarrow
\nu_e$ oscillations  with a mass scale $\Delta m^2\sim
1~\textrm{eV}^2$, perhaps ruling out a  \emph{light} sterile
neutrino. However, a recent analysis\cite{malto} suggests that while
  $(3+1)$ schemes are strongly disfavoured, $(3+2)$ neutrino schemes
  provide a good fit to both the LSND and MiniBooNE data, including the excess of low energy
  events, because of the possibility of CP violation in these schemes, although
  significant tension remains between appearance and disappearance experiments.

Sterile neutrinos as dark matter candidates would require masses in
the $\mathrm{keV}$
range\cite{dodelson,asaka,shi,kev1,hansen,kev2,kev3,kou,pastor,hannestad,bier},
and their radiative decay would contribute to the X-ray
background\cite{hansen,Xray}. Analysis from the X-ray background in
clusters provide constraints on the masses and mixing angles of
sterile neutrinos\cite{kou,boyarsky,hansen2,kou2}, and recently it
has been suggested that precision laboratory experiments  on $\beta$
decay in tritium   may be sensitive to $\sim \textrm{keV}$
neutrinos\cite{shapolast}.

Sterile neutrinos couple to standard model active neutrinos through
an off diagonal mass matrix, therefore they are produced via
active-sterile mixing. In the hot and dense environment of the early
Universe when the scattering rate of active neutrinos off the
thermal medium is large, namely a short mean free path, there is a
competition between the oscillation length and the mean free path.
It is expected  that when the oscillation length is much larger than
the mean free path, the active to sterile transition probability is
hindered because rapid scattering events ``freeze'' the state to the
active flavor state. This phenomenon receives the name of quantum
Zeno effect or Turing's paradox, studied early in quantum optical
coherence\cite{misra} but considered within the context of neutrino
oscillations in a medium in references\cite{stodolsky,sigl,raffkin}.
Pioneering work on the description of neutrino oscillations and
decoherence in a medium was cast in terms of kinetic equations for a
flavor ``matrix of densities''\cite{dolgov} or in terms of $2\times
2$ Bloch-type equations for flavor quantum mechanical
states\cite{stodolsky,enquist}. A general field theoretical approach
to neutrino mixing and kinetics was presented in
\cite{sigl,raffkin,raffeltbook,raffeltSN}, however, while such
approach in principle yields the time evolution of the distribution
functions, sterile neutrino production in the early Universe is
mostly studied in terms of simple phenomenological rate
equations\cite{dodelson,kev1,cline,kainu,foot,dibari}. An early
  approach\cite{cline} relied on a Wigner-Weisskopf effective
Hamiltonian for the quantum mechanical states in the medium, while
numerical studies of sterile neutrinos as possible dark matter
candidates\cite{kev1,dibari} rely on an approximate approach which
inputs an effective production rate in terms of a time averaged
transition probability\cite{kainu,foot} and relies on  the following
semiphenomenological rate
equation\cite{cline,foot,dibari,volkas,kev1}

\be \label{kine}\frac{d}{dt}f_s(p,t) \approx \Gamma(a\rightarrow
s;p) \left[f_a(p;t)-f_s(p;t)\right]  \ee where $f_{a,s}$ are the
distribution functions for active (a) and sterile (s) neutrinos,
$d/dt$ is the total time derivative including the redshift of
momenta through the expansion in the early Universe and
$\Gamma(a\rightarrow s;p)$ is an effective reaction rate. It is
determined to be\cite{cline,foot} \be
\label{effrate}\Gamma(a\rightarrow s;p) \approx \frac{\Gamma_{aa}
(p)}{2} \Big\langle P_{a\rightarrow s} \Big\rangle \ee where
$\Gamma_{aa} (p)\sim G^2_F \, p \,T^4$ is the active neutrino
reaction rate and $\Big\langle P_{a\rightarrow s} \Big\rangle$ is a
time average of the active-sterile transition probability in the
medium which in reference\cite{foot} is given by \be \Big\langle
P_{a\rightarrow s} \Big\rangle = \frac{\Gamma_{aa}}{2} \int^\infty_0
P_{a\rightarrow s}(t) dt \label{avedef}\ee with $P_{a\rightarrow s}$
the usual quantum mechanical  transition probability but
exponentially damped by a decoherence factor\cite{foot}\be
\label{Pasfoot} P_{a\rightarrow s}(t) =
\frac{\sin^22\theta_m}{2}~e^{-\frac{\Gamma_{aa}}{2}
t}~\left[1-\cos(\Delta E t)\right]\ee where $\Delta E,\theta_m$ are
the oscillation frequency and mixing angle in the medium
respectively and $\tau_{dec} = 2/\Gamma_{aa} $ is the decoherence
time scale. Hence the rate that enters in the kinetic equation
(\ref{kine}) is
  given by\cite{foot} \be  \Gamma(a\rightarrow s;p) =
\frac{\Gamma_{aa}(p)}{4}
\frac{\sin^22\theta_m(p)~\Big(\frac{2\,\Delta E(p)}{
\Gamma_{aa}(p)}\Big)^2}{\Big[1+\Big(\frac{2\,\Delta E(p)}{
\Gamma_{aa}(p)}\Big)^2\Big]} \label{prob}\ee where
$\theta_m(p),\Delta E(p)$ are the mixing angle and active-sterile
oscillation frequency in the medium respectively. The quantum Zeno
paradox is manifest in the ratio $2\Delta E(p)/\Gamma_{aa}(p)$ in
(\ref{prob}): for a relaxation time shorter than the oscillation
time scale, or mean free path smaller than the oscillation length,
$\Gamma_{aa}(p) \gg \Delta E(p)$ and the active-sterile transition
probability is suppressed, with a concomitant reduction of the
sterile production rate in the kinetic equation (\ref{kine}). Most
  studies\cite{kev1,dibari}  of the production of sterile neutrinos via
active-sterile mixing rely on the kinetic description afforded by
equation (\ref{kine}).

A field theoretic approach to   sterile neutrino production near a
MSW resonance which focuses primarily on the hadronic contribution
and seemingly yields a   different rate has   been proposed in
reference\cite{shaposhnikov06}, and more recently it has been
observed that quantum Zeno suppression may
 have  important consequences in thermal leptogenesis\cite{raffelt}.

{\bf Questions and goals:} Recently we have studied the
non-equilibrium aspects of oscillations and damping in a model of
mesons that effectively describes the dynamics of mixed neutrinos in
a medium in thermal equilibrium\cite{recentours}. In the case of two
species of ``neutrinos'' this study reveals that there are two
propagating modes in the medium, whose dispersion relations feature
the index of refraction correction from forward scattering similar
to those for neutrinos in the medium but also \emph{two different}
damping rates which are determined by the imaginary part of the
self-energy correction evaluated on the mass shell. For the case of
two mixed neutrinos in a medium it is natural to expect that the
imaginary part of the self-energy corrections evaluated on the mass
shell (dispersion relations) yield \emph{two different} damping
rates. Thus the results of ref.\cite{recentours} lead us to expect
that for one active and one sterile mixed neutrinos  the propagating
modes in the medium feature \emph{two different damping rates} and
this observation motivates the \textbf{first question}: How does the
active-sterile transition probability $P_{a\rightarrow s}(t)$
account for two damping scales?, namely why does the result
(\ref{Pasfoot}) feature only one damping scale?. This   question
leads to the related \textbf{second question}: How to generalize the
concept of a transition probability to the case of propagation in a
\emph{medium}?. The usual transition probability is based on the
evolution of single particle quantum mechanical wavefunctions which
are linear superpositions of the eigenstates of the Hamiltonian. The
statistical description of a  \emph{medium} does not rely on single
particle wavefunctions or quantum mechanical states but on the
\emph{quantum density matrix}. Therefore the concept of the
active-sterile transition probability $P_{a\rightarrow s}(t)$ in a
medium must be generalized in terms of the quantum density matrix.
As discussed above the final expression for the effective sterile
production rate (\ref{prob}) exhibits the Quantum Zeno suppression
whenever $\Gamma_{aa}(p) \gg \Delta E(p)$ which has been argued to
be the case at high temperature\cite{kev1}. From the quantum field
theory perspective this possibility is puzzling for the following
reason: at high temperature the difference in the oscillation
frequencies $\Delta E(p)$ is determined by the index of refraction
correction from forward scattering\cite{notzold} in the medium. This
is determined by a one-loop contribution\cite{notzold} and is
formally of order $G_F$, whereas the interaction rate $\Gamma_{aa}$
arises from an absorptive part of the self-energy and in the
effective Fermi's field theory is at least of two-loop order,
formally of order $G^2_F$. Therefore from the field theoretical
perspective quantum Zeno suppression requires a competition of terms
of different order in the perturbative expansion in Fermi's
effective field theory\cite{charged}. This observation brings us to
the \textbf{third question}: considering and active neutrino with
standard model interactions, can quantum Zeno suppression be
manifest at high temperatures within the regime of validity of the
perturbative expansion?.

 The emerging cosmological and
astrophysical importance of sterile neutrinos motivates a deeper
scrutiny of the current approach to the dynamical aspects of their
production based on the rate equation (\ref{kine}) with the
effective rate(\ref{prob}).   While at this stage we take this
description based on (\ref{kine}-\ref{avedef}) for \emph{granted},
our goal is to address the three questions enunciated above within
the quantum field theory of mixed neutrinos with standard model
interactions in the medium, and in so doing we scrutinize the
reliability of this approach. Our goals in this article are: {\bf
i:} to provide a quantum field theoretical understanding of the
dispersion relations and  damping rates of the two propagating modes
(quasiparticles) in the medium, {\bf ii:} to provide a
generalization of the active-sterile transition probability in
\emph{real time} in the medium and a reassessment of the time
averaged transition probability $ \langle P_{a\rightarrow s}
\rangle$ directly from the non-equilibrium time evolution of the
\emph{full density matrix} and {\bf iii:} to scrutinize the
possibility of  quantum Zeno suppression within the realm of
validity of perturbation theory with standard model interactions for
the active neutrino.

{\bf Main results:}

We consider   one active and one sterile neutrino\cite{fuller3} to
highlight the main conceptual aspects. Unlike most treatments in the
literature that study the dynamics in terms of Bloch-type equations
for  a $2\times 2$ flavor density
matrix\cite{stodolsky,foot,dolgovrev}, we study the \emph{full
quantum field theoretical density matrix}. A main advantage of
studying the time evolution of the   density matrix directly within
the   quantum  field theory context  is that we obtain the
\emph{neutrino propagator} which yields the quasiparticle dispersion
relations and damping rates in the medium. Furthermore, the time
evolution of the quantum field density matrix allows to study the
non-equilibrium dynamics of neutrino mixing and propagation as an
initial value problem from which we obtain the dispersion relation
of the correct quasiparticle modes in the medium, their damping
rates (widths) and the generalization of the active-sterile
transition probability. These are all determined by the neutrino
propagator \emph{in the medium} which includes self-energy
corrections up to two loops $\mathcal{O}(G^2_F)$ in the standard
model weak interactions. Our main results are the following:
\begin{itemize}
\item{  There are two  quasiparticle propagating modes, their dispersion relations are the usual ones in
terms of the index of refraction in the medium\cite{notzold} plus
perturbative radiative corrections of $\mathcal{O}(G^2_F)$  and
their damping rates are given by $\Gamma_1(p) = \Gamma_{aa}(p)
\cos^2\theta_m(p)~ ; ~\Gamma_2(p) = \Gamma_{aa}(p)
\sin^2\theta_m(p)$ where $\Gamma_{aa}(p)\propto G^2_F\,p \,T^4$ is
the active neutrino scattering rate in the absence of mixing, and
$\theta_m(p)$ the mixing angle in the medium. We provide a physical
interpretation of these different quasiparticle relaxation rates and
argue that these must naturally be correct in agreement with the
fact that sterile neutrinos are much more weakly coupled to the
plasma than active neutrinos in the  regimes far away from an MSW
resonance. }

\item{We generalize the concept of the active-sterile transition
probability $P_{a\rightarrow s}(t)$ \emph{in the medium} from
expectation values of the active and sterile neutrino field
operators in the \emph{full quantum density matrix}. This is
achieved by furnishing an initial value problem via linear response:
the density matrix is initialized to feature an non-vanishing
expectation value of the active neutrino field, but a vanishing
expectation value of the sterile field. Upon time evolution a
non-vanishing expectation value of the sterile neutrino field
develops from which we extract unambiguously the transition
probability. This formulation directly inputs the neutrino
propagator with self-energy corrections up to $\mathcal{O}(G^2_F)$.
We find \be \label{Pintrodu} P_{a\rightarrow s}(t) =
\frac{\sin^22\theta_m}{4}\left[e^{-\Gamma_1t} + e^{-\Gamma_2
t}-2e^{-\frac{1}{2}(\Gamma_1+\Gamma_2)t}~\cos\big(\Delta E
t\big)\right]  \ee This expression identifies the decoherence time
scale for suppression of the interference term $\tau_{dec} =
2/(\Gamma_1+\Gamma_2)=2/\Gamma_{aa}$. Although the oscillatory
interference term in the active-sterile transition probability is
suppressed on the \emph{decoherence} time scale $\tau_{dec}$, far
away from an MSW resonance  the relevant time scale for suppression
of $P_{a\rightarrow s}(t)$ is determined by the \emph{smaller} of
the relaxation scales for the quasiparticles $\Gamma_1(p)$ or
$\Gamma_2(p)$. \emph{If} the effective sterile production rate
(\ref{effrate}) is computed by inserting the result (\ref{Pintrodu})
into the time average (\ref{avedef}) the result for the effective
production rate is \emph{enhanced} far away from MSW resonances
compared to that given by (\ref{prob}). However we argue that the
widely different damping rates $\Gamma_{1,2}$ \emph{suggest a
breakdown of the simple rate equation} (\ref{kine}) in these regimes
in the early Universe. }

\item{

We   provide a   \emph{real time} interpretation of the quantum Zeno
suppression based on the generalization of the active-sterile
transition probability (\ref{Pintrodu}). The \emph{complete general
conditions} for
       quantum Zeno suppression of the active-sterile transition probability are found to be:
\begin{itemize} \item{{\bf a) }
        the active neutrino scattering rate much larger than the oscillation frequency
        $\Gamma_{aa}(p) \gg \Delta E(p)$,} \item{{\bf b) }
         The relaxation rates of the propagating modes must be approximately equal. In the
        case under consideration with  $\Gamma_1(p) =
\Gamma_{aa}(p) \cos^2\theta_m(p) ~;~ \Gamma_2(p) = \Gamma_{aa}(p)
\sin^2\theta_m(p)$ this condition determines an MSW resonance in the
medium.} \end{itemize} Although these conditions are general, for
sterile neutrinos with $m_s \sim \textrm{keV}$ and $\sin2\theta
\lesssim
        10^{-3}$ and   standard model interactions for the active neutrino,
        we find that they   \emph{may only be}
          fulfilled   near an MSW resonance at $T_{MSW}
        \sim 215 \,\textrm{MeV}$, but a firm assessment of such possibility requires to include $\mathcal{O}(G^2_F)$
        corrections to the index of refraction and a deeper assessment of the perturbative expansion. Far away from the resonance either
        at high or low temperature there is a wide separation between the two relaxation rates of the propagating modes
        in the medium. In these cases the transition probability
        \emph{reaches a maximum} on the \emph{decoherence} time
        scale $\tau_{dec}=2/\Gamma_{aa}$,  and is suppressed on a \emph{much
        longer time scale} determined by the \emph{smaller} of the
        damping rates.
Even for $ \tau_{dec}\Delta E(k)\ll 1 $, which in the literature
\cite{stodolsky,kev1} is taken to indicate  quantum Zeno
suppression, we find that        the transition probability is
substantial on time scales
     much longer than $\tau_{dec}$ if $\Gamma_1$ and $\Gamma_2$ are \emph{widely
     separated}.
  }

\end{itemize}

  Section
       (\ref{sec:eqnofmot}) provides  a   study of
        the time evolution of the \emph{full quantum field
       theory density matrix},  and the equations of motion for expectation values of the
       neutrino fields. In this section we obtain the
       dispersion relations and widths of the propagating modes
       (quasiparticles) in the medium up to second order in the weak
       interactions. In section (\ref{sec:QZE}) we introduced the
       \emph{generalized} transition probability in the medium from
       the time evolution of expectation values of neutrino field
       operators in the density matrix.   In this section   we discuss
       in detail the conditions for the quantum Zeno effect, both in real time and in the time-averaged
       transition probability and the possibility of quantum Zeno suppression within the realm of validity
       of perturbation theory in standard model weak interactions. In
       section (\ref{sec:implications}) we discuss the implications
       of our results for the production of sterile neutrinos in the
       early Universe. In this section we argue that the in the early Universe far away from an MSW resonance the
       wide separation of the damping scales   makes \emph{any}  definition of
       the time averaged transition probability \emph{ambiguous},
       and question the validity of the usual rate equation to describe sterile
       neutrino production in the early Universe far away from MSW resonances.  Section
       (\ref{sec:conclu}) presents our conclusions .

\section{Quantum field theory treatment in the medium}\label{sec:eqnofmot}
\subsection{ Non-equilibrium density matrix}
In a medium the relevant question is \emph{not} that of the time
evolution of a pure quantum state, but more generally that of a
density matrix from which expectation values of suitable operators
can be obtained.

In order to provide a detailed understanding of the quantum Zeno
effect, we need a reliable estimate of the dispersion relations
  and the damping rates of the propagating modes in the
medium which are determined by the complex poles of the neutrino
propagator  in the medium.

  In this article we obtain these from the study of the real
time evolution of the \emph{full density matrix}  by implementing
the methods of quantum field theory in real time described in
references\cite{schwinger,maha,keldysh,revi,eunu,recentours}. This
is achieved by introducing external (Grassmann) sources that induce
an expectation value for the neutrino fields. Upon switching off the
sources these expectation values relax towards equilibrium and their
time evolution reveals both the correct energy and the relaxation
rates\cite{eunu,recentours}. The main ingredient in this program is
the active neutrino self-energy which we obtain up to second order
in the standard model weak interactions.

We consider a model of one active and one sterile Dirac neutrinos in
which active-sterile mixing is included via an off diagonal Dirac
mass matrix and the active neutrino only features standard model
weak interactions. The relevant Lagrangian density is given by

\be \mathcal{L} = \mathcal{L}^0_\nu + \mathcal{L}_{Ia} \,,\ee where

 \be\label{FFnuL} \mathcal{L}^0_{\nu} = \overline{\nu}
\left(i {\not\!{\partial}}\, \mathbb{I}-\mathbb{M}\right) \nu \,,\ee
with  $\nu$ being the neutrino doublet

\be \nu \equiv     \left(
         \begin{array}{c}
           \nu_{a} \\
           \nu_{s  } \\
         \end{array}
       \right) \,,\ee

\noindent and $  a,s$ refer to the flavor indexes of the active and
sterile neutrinos respectively.


The neutrino fields are four component Dirac spinors
$\nu=\nu_R+\nu_L$ with $R,L$ the right and left handed components,
both for $a,s$. The mass matrix in (\ref{FFnuL}) is of the Dirac
type: $\overline{\nu_R}\, \mathbb{M}\, \nu_L + \mathrm{h.c.}$.

 The Dirac  mass matrix $\mathbb{M}$
is given by

\be \label{massmatrix} \mathbb{M}=\left(%
\begin{array}{cc}
  m_{aa} & m_{as} \\
  m_{as} & m_{ss} \\
\end{array}%
\right)\,. \ee It can be diagonalized by the unitary transformation
that takes flavor into mass eigenstates, namely \be  \left(
         \begin{array}{c}
           \nu_{a} \\
           \nu_{s  } \\
         \end{array}
       \right) = U(\theta)  \left(
         \begin{array}{c}
           \nu_{1} \\
           \nu_{2  } \\
         \end{array}
       \right) \,,\label{unitrafo} \ee
 with the unitary transformation given by the $2\times 2$ matrix
\be \label{vacrot} U(\theta) = \left(
          \begin{array}{cc}
            \cos\theta  & \sin \theta  \\
            -\sin \theta  & \cos \theta  \\
          \end{array}
        \right)\,. \ee In this basis the mass matrix is diagonal  \be \mathbb{M} = \left(
                    \begin{array}{cc}
                      M_1 & 0 \\
                      0 & M_2 \\
                    \end{array}
                  \right) \,,\label{massdiag}\ee with the relation \be m_{aa} = \cos^2\theta M_1+\sin^2\theta
M_2~~;~~m_{ss}=\sin^2\theta M_1 + \cos^2\theta M_2
~~;~~m_{as}=\frac{1}{2}(M_2-M_1) \sin2\theta \,,\label{masses}\ee
where
  $\theta$ is the \emph{vacuum} mixing angle. The
  Lagrangian density $\mathcal{L}_{Ia}$ describes the weak
interactions of the active neutrino $\nu_a$ with hadrons or quarks
and its associated charged lepton. Leptons, hadrons or quarks reach
equilibrium in a thermal bath on time scales far shorter than those
of neutrinos, therefore in what follows we assume these degrees of
freedom to be in thermal equilibrium. Furthermore, in our analysis
we will \emph{not} include the non-linearities associated with a
neutrino background, such component requires a full non-equilibrium
treatment and is not germane to the focus of this study. The
Lagrangian density that includes both charged and neutral current
interactions can be written in the
form\cite{raffkin,raffeltSN,raffeltbook}

\be \label{lagraIa}\mathcal{L}_{Ia} =  \frac{G_F}{\sqrt{2}}~\left[
\overline{\mathcal{O}}_a L \nu_a + \mathrm{h.c.}+  \overline{\nu}_a
\gamma_\mu J^\mu L \nu_a \right]\ee where $L= (1-\gamma^5)/2$, the
current $J^\mu$ includes both charge and neutral current
contributions from the background in thermal equilibrium. For
example the following charged and neutral current contributions to
the effective Lagrangian (taking the active to be, for example, the
electron neutrino): \be
-\frac{G_F}{\sqrt{2}}\Bigg\{4[\overline{\nu}_e \gamma^\mu L
e][\overline{e}\gamma_{\mu}L \nu_e]+ 2\overline{\nu}_e \gamma^\mu
[\overline{e}\gamma_\mu (g_V-g_A \gamma^5)e]L \nu_e \Bigg\}\,, \ee
the first term (from charged currents) can be Fierz-rearranged to
yield the form of the second term in (\ref{lagraIa}). This term
yields a contribution to the index of refraction in the
medium\cite{notzold}, $\mathcal{O}_a$ describes the charged current
interaction with hadrons or quarks and the charged lepton, for
example\cite{raffkin,raffeltbook} $\mathcal{O}_a = \gamma^\mu L
\psi_e \overline{\psi}_n \gamma_\mu (C_V-C_A\gamma^5)\psi_p$.  In
the case of all active species the neutral current contribution to
$J^\mu$ is the same for all flavors (when the neutrino background is
neglected), hence it does not contribute to oscillations and the
effective matter potential. In the case in which there are sterile
neutrinos, which do not interact with the background directly, the
neutral current contribution does contribute to the medium
modifications of active-sterile mixing angles and oscillations
frequencies.

To study the dynamics in a medium we must consider the time
evolution of the \emph{density matrix}. While the usual approach
truncates the full density matrix to a $2\times 2$ ``flavor''
subspace thus neglecting all but the flavor degrees of freedom, and
studies its time evolution in terms of Bloch-type
equations\cite{stodolsky,foot}, our study relies instead on the time
evolution of the \emph{full quantum field theoretical density
matrix}.


The full density matrix describes a statistical ensemble of
neutrinos \emph{and} charged leptons, quarks or hadrons, these
latter degrees of freedom    are in thermal equilibrium and
constitute the thermal bath. The fact that the density matrix
describes charged leptons, quarks and or hadrons in statistical
equilibrium will be used below (see eqns.  \ref{sigtad},
\ref{sigretGF2}  ) when the correlation functions of these fields
are obtained from ensemble averages in the density matrix.


The time evolution of the quantum density matrix $\hat{\rho}$ is
given by the quantum Liouville equation \be i
\frac{d\hat{\rho}(t)}{dt} =
\left[H,\hat{\rho}(t)\right]\label{liouville}\ee where $H$ is the
full Hamiltonian with weak interactions. The solution is given by
\be \hat{\rho}(t) = e^{-iHt}\,\hat{\rho}(0) \, e^{ iHt}
\label{solul}\ee from which the time evolution of observables
associated with an operator $\mathcal{A}$, namely its expectation
value in the time evolved density matrix is given by \be \langle
\mathcal{A}(t) \rangle = \mathrm{Tr}\hat{\rho}(t) \mathcal{A}  \,.
\label{expe} \ee

  The density matrix elements  in the field basis
are given by \be \hat{\rho}(\psi,\psi';t) = \int \mathcal{D}\phi
\mathcal{D}\phi'\, \langle \psi| e^{-iHt} |\phi \rangle \,
\hat{\rho}(\phi,\phi';0)\, \langle \phi'| e^{ iHt} |\psi' \rangle\,,
\ee the matrix elements of the forward and backward time evolution
operators can be handily written as path integrals and the resulting
expression involves a path integral along a forward and backward
contour in time. This is the
Schwinger-Keldysh\cite{schwinger,keldysh,maha,revi}  formulation of
non-equilibrium quantum statistical mechanics which yields the
correct time evolution of quantum density matrices in field theory.
Expectation values of operators are obtained as usual by coupling
sources conjugate to these operators in the Lagrangian and taking
variational derivatives with respect to these sources. This
formulation of non-equilibrium quantum field theory   yields all the
correlation and Green's functions. Of primary focus  is the
\emph{neutrino retarded propagator} \be S_{\alpha
\beta}(\vec{x}-\vec{x}';t-t') = -i \Theta(t-t')~
 \mathrm{Tr}\hat{\rho}(0)
\big\{\psi_{\alpha}(\vec{x},t),\overline{\psi}_{\beta}(\vec{x'},t')
\big\}  \,, \label{propa1}\ee where the flavor indices
$\alpha,\beta$ correspond to either active or sterile and
$\psi_\alpha(\vec{x},t)$ is a neutrino field in the Heisenberg
picture. The (complex) poles in complex frequency space of the
spatio-temporal Fourier transform of the neutrino propagator yields
the dispersion relations and damping rates of the quasiparticle
states in the medium. It is not clear if this important information
can be extracted from the truncated $2\times 2$ density matrix in
flavor space usually invoked in the literature and which forms the
basis of the kinetic description (\ref{kine}), but certainly the
\emph{full quantum field density matrix} does have all the
information on the correct dispersion relations and relaxation
rates.

 A standard
approach to obtain the propagation frequencies and damping rates of
quasiparticle excitations in a medium is the method of linear
response\cite{fetter}. An external source $\eta$ is coupled to the
field operator $\psi$ to induce an expectation value of this
operator in the many body density matrix, the time evolution of this
expectation value yields the quasiparticle dynamics, namely the
propagation frequencies and damping rates. In linear response \be
\langle \psi_{\alpha}(\vec{x},t)\rangle \equiv
\mathrm{Tr}~\hat{\rho}(0) \psi_{\alpha}(\vec{x},t)   = -\int d^3x'
dt' S_{\alpha \beta}(\vec{x}-\vec{x}';t-t')\,
\eta_{\beta}(\vec{x}',t')\,,\label{lr} \ee where
$S(\vec{x}-\vec{x}';t-t') $ is the retarded propagator or Green's
function (\ref{propa1}) and averages are in the full quantum density
matrix. The quasiparticle dispersion relations and damping rates are
obtained from the complex poles of the spatio-temporal Fourier
transform of the retarded propagator in the complex frequency
plane\cite{fetter,ftf}. For one active and one sterile neutrino
there are two propagating modes in the medium. Up to one loop order
$\mathcal{O}(G_F)$ the index of refraction in the medium yields two
different dispersion relations\cite{notzold}, hence we \emph{expect}
also that the damping rates for these two propagating modes which
will be obtained up to $\mathcal{O}(G^2_F)$ will be different. This
expectation will be confirmed below with the explicit computation of
the propagator up to $\mathcal{O}(G^2_F)$.

Linear response is the standard method to obtain the dispersion
relations and damping rates of quasiparticle excitations in a plasma
in finite temperature field theory\cite{ftf}.  The linear response
relation (\ref{lr}) can be inverted to write \be S^{-1} \langle
\psi(\vec{x},t) \rangle = -\eta(\vec{x},t)\,, \label{eqm}\ee where
$S^{-1}$ is the (non-local) differential operator which is the
inverse of the propagator, namely the \emph{effective Dirac
operator} in the medium that includes self-energy corrections. This
allows to study the dynamics as an \emph{initial value problem} and
to recognize the quasiparticle dispersion relations and damping
rates directly from the time evolution of expectation values of the
field operators. This method has been applied to several different
problems in quantum field theory out of equilibrium  and the reader
is referred to the literature for detailed
discussions\cite{disip,tadpole,nosfermions,eunu,recentours}.

It is important to highlight that $\langle \psi(\vec{x},t) \rangle$
\emph{is not} a single particle wave function but an ensemble
average of the quantum field operator in the non-equilibrium density
matrix, namely an ensemble average. In contrast to this expectation
value, a single particle wave function is defined as $\langle
1|\psi(\vec{x},t)|0\rangle$ where $|0\rangle$ is the vacuum and
$|1\rangle$ a Fock state with \emph{one} single particle.

In the present case the initial value problem allows us also to
study the time evolution of flavor off diagonal density matrix
elements. Consider an external source $\eta_a$ that induces an
initial expectation value only for the active neutrino field
$\psi_a$, such an external source \emph{prepares} the initial
density matrix so that at $t=0$ the active neutrino field operator
features a non-vanishing  expectation value, while the sterile one
has a vanishing expectation value. Upon time evolution the density
matrix develops \emph{flavor off diagonal matrix elements} and the
\emph{sterile} neutrino field $\psi_s$ develops an expectation
value. The solution of the equation of motion (\ref{eqm}) as an
initial value problem allows us to extract precisely the time
evolution of $\langle \psi_s \rangle$ from which we unambiguously
extract the transition probability in the medium.


\subsection{Equations of motion in linear response}

The linear response approach to studying the non-equilibrium
evolution relies on ``adiabatically switching on'' an external
source $\eta$ that initializes the quantum density matrix to yield
an expectation value for the neutrino field(s). Upon switching off
the external source the expectation values of the neutrino fields
relax to equilibrium. The real time evolution of the expectation
values reveals the dispersion relations and damping rates  of the
propagating quasiparticle modes in the medium. These are determined
by the poles of the retarded propagator in the complex frequency
plane\cite{fetter,ftf}.

The equation of motion for the expectation value of the flavor
doublet is obtained by   introducing   external Grassmann-valued
sources $\eta$\cite{disip,tadpole,nosfermions}

\be \mathcal{L}_S = \overline{\nu} \, \eta  + \overline{\eta} \, \nu
\, , \label{Lsource} \ee  shifting the field

\be \nu^{\pm}_\alpha = \psi_\alpha + \Psi^{\pm}_\alpha ~~;~~
\psi_\alpha = \langle \nu^{\pm}_\alpha \rangle ~~;~~ \langle
\Psi^{\pm}_\alpha \rangle =0, \label{shift}\ee for $\alpha=a,s$, and
imposing $\langle \Psi^{\pm}_a \rangle =0$ order by order in the
perturbation theory \cite{disip,tadpole,nosfermions}. Implementing
this program   we find the following equation of motion for the
expectation value of the neutrino field $\psi_{\alpha}(\vx,t)=
\mathrm{Tr}\big[\hat{\rho}(0)\nu_{\alpha}(\vx,t)\big]$ induced by
the external field $\eta$

\be
\left(i\not\!{\partial}\,\delta_{\alpha\beta}-M_{\alpha\beta}+\Sigma^{tad}_{\alpha\beta}L\right)\,\psi_\beta
(\vx,t) + \int d^3 x' \int^t_{-\infty} dt'
\Sigma^{ret}_{\alpha\beta}(\vx-\vx',t-t')\psi_\beta(\vx',t') = -
\eta_\alpha(\vx,t), \label{eqnofmot}\ee The tadpole \be \Sigma^{tad}
= \frac{G_F}{\sqrt{2}} \gamma_0 \,\mathrm{Tr}\widehat{\rho}(0)\,
J^0(0,0) \label{sigtad}\ee describes the one-loop charged and
neutral current contributions to the matter potential in the medium,
and \be\Sigma^{ret}_{aa}(\vec{x}-\vec{x'};t-t') =
\frac{i\,G^2_F}{2}~\mathrm{Tr}\hat{\rho}(0)\Big[\mathcal{O}_a(\vec{x},t)\overline{\mathcal{O}}_a(\vec{x'},t')+
\overline{\mathcal{O}}_a(\vec{x'},t')\mathcal{O}_a(\vec{x},t)\Big]\,.\label{sigretGF2}
\ee The latter describes the two-loops diagrams with intermediate
states of hadrons or quarks and the charged lepton, it is a
\emph{fermionic} correlation function in equilibrium  and its
spatio-temporal Fourier transform features an imaginary part that
yields the relaxation rates of neutrinos in the medium. As shown in
ref.\cite{eunu}, the spatial Fourier transform of the retarded
self-energy can be written as

\be \Sigma^{ret}(\vk,t-t') =   \, \frac{i}{\pi}
\int_{-\infty}^{\infty} dk_0 \,
\mathrm{Im}\Sigma(\vk,k_0)\,e^{ik_0(t-t')} \,. \label{sigreta}\ee

The  imaginary part $\mathrm{Im}\Sigma(\vk,k_0)$ evaluated on the
mass shell of the propagating modes determines the relaxation rate
of the neutrinos in the medium. Since only the active neutrino
interacts with the degrees of freedom in the medium, both
self-energy contributions are of the form

\be \Sigma = \left(
               \begin{array}{cc}
                 \Sigma_{aa} & 0 \\
                 0 & 0 \\
               \end{array}
             \right) \,.\label{selfenerform} \ee

The   \emph{initial value problem} is set up as follows\cite{eunu}.
Consider an external Grassman valued source adiabatically switched
on at $t=-\infty$ and off at $t=0$, \be \eta_\alpha(\vx,t) =
\eta_\alpha(\vx,0)~e^{\epsilon t}~\theta(-t)~~\epsilon \rightarrow
0^+ \,. \label{adiaeta}\ee It is straightforward to confirm that the
solution of the equation of motion (\ref{eqnofmot}) for $t<0$ is
given by \be \psi_\beta(\vx,t) = \psi_\beta(\vx,0)~e^{\epsilon
t}\,.\label{negtsol}\ee Inserting this solution for $t<0$ the
equation of motion determines a relation between
$\psi_{\beta}(\vx,0)$ and $\eta_{\alpha}(\vx,0)$. For $t>0$ the
equation of motion becomes an \emph{initial value problem} with
initial value given by  $\psi_{\beta}(\vx,0)$. For $t>0$ introducing
spatial Fourier transforms and taking the Laplace transform,  the
equation of motion   becomes (see   ref.\cite{eunu} for details)

\be \left[ \left( i\gamma^0 s-\vec{\gamma}\cdot \vec{k}  \right)
\,\delta_{\alpha\beta}-M_{\alpha\beta}+\Sigma^{tad}_{\alpha\beta}\,L
+ \widetilde{\Sigma}_{\alpha\beta}(\vk,s)\,L
\right]\widetilde{\psi}_\beta(\vk,s) =
i\left(\gamma^0\,\delta_{\alpha\beta}+\mathcal{O}(G^2_F)\right)\psi_\beta(\vk,0)\,.
\label{lapladirac} \ee where $\widetilde{\psi},\widetilde{\Sigma}$
denote Laplace transforms with Laplace variable $s$. The Laplace
transform of the retarded self energy admits a dispersive
representation which follows from eqn.(\ref{sigreta}),
namely\cite{eunu}

\be \widetilde{\Sigma} (\vk,s) = \int^{\infty}_{-\infty}
\frac{dk_0}{\pi} \frac{\mathrm{Im}\Sigma(\vk,k_0)}{k_0-is}
\label{siglapla}\ee

Following ref.\cite{eunu}, we proceed to solve the equation of
motion by Laplace transform as befits an initial value problem.

In what follows we will ignore the perturbative corrections on the
right hand side of (\ref{lapladirac}) since these only amount  to a
perturbative multiplicative renormalization of the amplitude, (see
ref.\cite{eunu} for details).

The chiral nature of the interaction constrains the self-energy to
be of the form\cite{notzold,eunu}

\be \Sigma^{tad} \,L + \widetilde{\Sigma} (\vk,s)\,L =
\left(\gamma^0 \mathbb{A}(s ;k)-\vec{\gamma}\cdot \hat{\bf
k}\,\mathbb{B}(s;k)\right)L \ee where the matrices
$\mathbb{A},\mathbb{B}$ are of the form given in eqn.
(\ref{selfenerform}) with the only matrix elements being
$A_{aa};B_{aa}$ respectively. The dispersive form of the self-energy
(\ref{siglapla}) makes manifest that for $s$ near the imaginary axis
in the complex s-plane \be \widetilde{\Sigma} (\vk,s=-i\omega\pm
\epsilon) = \int^{\infty}_{-\infty} \frac{dk_0}{\pi}
\mathcal{P}\left[\frac{\mathrm{Im}\Sigma(\vk,k_0)}{k_0-\omega}\right]
\, \pm i\,\mathrm{Im}\Sigma(\vk,\omega)\,, \label{disc}\ee where
$\mathcal{P}$ indicates the principal part. This result will be
important below.

The solution of the algebraic matrix equation (\ref{lapladirac}) is
simplified by  expanding the left and right handed components of the
Dirac doublet $\widetilde{\psi}$ in the helicity basis as

\be \widetilde{\psi}_L = \sum_{h=\pm1} \left(
                        \begin{array}{ c}
                            0 \\
                          v^{(h)}\otimes \widetilde{\varphi}^{(h)} \\
                        \end{array}
                      \right)  ~~;~~ \widetilde{\psi}_R = \sum_{h=\pm1} \left(
                        \begin{array}{ c}
                            v^{(h)}\otimes \widetilde{\xi}^{(h)} \\
                          0 \\
                        \end{array}
                      \right) \ee where the Weyl spinors $ v^{(h)}$
                      are eigenstates of   helicity  $\vec{\sigma}\cdot \hat{\bf k}$  with
                      eigenvalues $h=\pm 1$ and $\widetilde{\varphi}^{(h)};\widetilde{\xi}^{(h)}$ are flavor doublets with the
upper component being the active and the lower the sterile
neutrinos.

Projecting the  equation of motion (\ref{lapladirac}) onto right and
left handed components and onto helicity eigenstates, we find after
straightforward algebra

\be \left[-(s^2+k^2)\mathbb{I}+(is-h k)( {\mathbb{A}(k;s)}+h \,
\mathbb{B}(k;s))-\mathbb{M}^2 \right]\widetilde{\varphi}^{(h)}
(\vk,s)= i(is-h k) \mathbb{I}\,\varphi^{(h)} (\vk,0) -i \mathbb{M}
\,\xi^{(h)}(\vk,0) \label{lapLproy} \ee

\be\widetilde{\xi}^{(h)} (\vk,s) =
-\frac{is+hk}{s^2+k^2}\left[-\mathbb{M}\,
\widetilde{{\varphi}}^{(h)} (\vk,s)+i  \xi^{(h)}
(\vk,0)\right]\label{xitil} \ee where again we have neglected
perturbatively small corrections on the right hand side of eqn.
(\ref{lapLproy}).

It proves convenient to introduce the following definitions,

\bea && \delta M^2   =   M^2_1-M^2_2 ~~;~~ \overline{M}^{\,2} = \frac{1}{2}(M^2_1+M^2_2)\label{deltaM}\\
&& S_h(k;s)   = (is-h k)(A_{aa}(k;s)+h B_{aa}(k;s)) \label{Sh} \\
&& \Delta_h(k;s)= \frac{S_h(k;s)}{\delta M^2}\label{Deltah}\\&&
\rho_h(k;s) = \left[\left(\cos2\theta - \Delta_h(k;s)\right)^2 +
\sin^22\theta\right]^{\frac{1}{2}} \label{rho}\\ &&
\cos2\theta^{(h)}_m(k;s)   =   \frac{\cos2\theta
-\Delta_h(k;s)}{\rho_h(k;s)} \label{cosmed}
\\&& \sin 2\theta^{(h)}_m(k;s)   =   \frac{\sin2\theta }{\rho_h(k;s)}\eea in
terms of which

\bea \label{matrix} && -(s^2+k^2)\mathbb{I}+(is-h k)(
{\mathbb{A}(k;s)}+h \, \mathbb{B}(k;s))-\mathbb{M}^2     =
 \left(-s^2-k^2+\frac{1}{2}S_h(k,s)-\overline{M}^{\,2}\right)
\mathbb{I} \nonumber \\&& - \frac{\delta M^2}{2}\,\rho_h(k;s)\left(
          \begin{array}{cc}
           \cos 2\theta^{(h)}_m(k;s)&  - \sin 2\theta^{(h)}_m(k;s)  \\
             - \sin 2\theta^{(h)}_m(k;s) &  - \cos 2\theta^{(h)}_m(k;s)  \\
          \end{array} \right) \eea

The solution of the equation (\ref{lapLproy}) is given by

 \be \widetilde{\varphi}^{(h)} (\vk,s)=   \widetilde{\mathbb{S}}^{(h)}(k,s)\,\Bigg[-i \mathbb{M} \,\xi^{(h)}(\vk,0)+i(is-h k)
\mathbb{I}\,\varphi^{(h)} (\vk,0) \Bigg] \label{laplasol} \ee where
the propagator $\widetilde{\mathbb{S}}^{(h)}(k,s)$ is given by

\be \label{propa} \widetilde{\mathbb{S}}^{ (h) }(k,s) =
\frac{1}{\left[\alpha^2_{h}(s,k)-\beta^2_{h}(s,k)\right]} \Bigg[
\alpha_{h}(s,k) \, \mathbb{I} + \beta_h(s,k) \left(
          \begin{array}{cc}
           \cos 2\theta^{(h)}_m(k;s)&  - \sin 2\theta^{(h)}_m(k;s)  \\
             - \sin 2\theta^{(h)}_m(k;s) &  - \cos 2\theta^{(h)}_m(k;s)  \\
          \end{array} \right)
        \Bigg] \ee and we defined

\bea  \alpha_{h}(k;s) &  = &  \Bigg[ -(s^2 + k^2)
+\frac{1}{2} \; S_h(k;s)-\overline{M}^{\,2}\Bigg] \label{alfa1}\\
\beta_{h}(k;s)  & = &     \frac{\delta M^2}{2}\;   \rho_{h}(k;s) \;.
\label{beta1} \eea

The real time evolution is obtained by inverse Laplace transform,

\be \varphi^{(h)}(\vk,t) = \int_{\Gamma} \frac{ds}{2\pi i}\;
\widetilde{\varphi}^{(h)}(\vk,s)\; e^{st}   \; ,\label{inverlapla}
\ee \noindent where $\Gamma$ is the Bromwich contour in the complex
$s$ plane running parallel to the imaginary axis to the right of all
the singularities of the function $\widetilde{\varphi}(\vk,s)$ and
closing on a large semicircle to the left of the imaginary axis. The
singularities of $\widetilde{\varphi}(\vk,s)$ are those of the
propagator (\ref{propa}). \emph{If the particles are asymptotic
states and do not decay} these are isolated simple poles along the
imaginary axis away from multiparticle cuts. However, in a medium or
for decaying states, the isolated poles move into the continuum of
the multiparticle cuts and off the imaginary axis. This is the
general case of resonances which correspond to poles in the second
or higher Riemann sheet and the propagator is a complex function
with a branch cut along the imaginary axis in the complex s-plane as
indicated by eqn. (\ref{disc}). Its analytic continuation onto the
physical sheet features the usual Breit-Wigner resonance form and a
complex pole and the width determines the damping rate of
quasiparticle excitations\cite{disip,tadpole,nosfermions}.


It is important and relevant to highlight that the full width or
damping rate is the \emph{sum} of all the partial widths that
contribute to the damping from different physical processes: decay
if there are available decay channels, and in a medium the
collisional width and or Landau damping also contribute to the
imaginary part of the self-energy on the mass shell. The
quasiparticle damping rate is one-half the relaxation rate in the
Boltzmann equation for the distribution
functions\cite{weldon,nosfermions}.

It is convenient to change the integration variable to $s=-i\omega +
\epsilon$ with $\epsilon \rightarrow 0^+$ and to write the real time
solution (\ref{inverlapla}) as follows

\be \varphi^{(h)}(\vk,t) = \int_{-\infty}^{\infty}
\frac{d\omega}{2\pi }\;
\widetilde{\varphi}^{(h)}(\vk,s=-i\omega+\epsilon)\; e^{-i\omega
\,t} \; ,\label{inver} \ee

We focus on   ultrarelativistic neutrinos which is the relevant case
in the early Universe. Let us consider that initially there are no
right handed neutrinos and only negative helicity are produced,
namely $h=-1$, and denoting the negative helicity doublet of
expectation values as

\be \varphi^{(-1)} (\vk,t) = \left(  \begin{array}{c}
                                       \nu_a(\vk,t) \\
                                       \nu_s(\vk,t)
                                     \end{array} \right)  \,,
                                     \label{inifi}\ee where $\nu_{a,s}(\vk,t)$ now represent the
                                     \emph{expectation values} of the negative helicity components of the
                                     neutrino fields in the   density matrix. We find
                                     the expectation values at time
                                     $t$   given by

\be \left(\begin{array}{c}
      \nu_a(\vk,t) \\
      \nu_s(\vk,t)
    \end{array} \right) = i \int_{-\infty}^{\infty}
\frac{d\omega}{2\pi }\; e^{-i\omega\,t}\, (\omega + k)\, G(k;\omega)
~~ \left(\begin{array}{c}
      \nu_a(\vk,0) \\
      \nu_s(\vk,0)
    \end{array} \right) \label{nusoft} \ee \noindent where

    \be G(k;\omega) \equiv \widetilde{\mathbb{S}}^{ (-1)
    }(k,s=-i\omega+\epsilon) \label{propaw}\ee and the integral in
    (\ref{nusoft}) is carried out in the complex $\omega$ plane
    closing along a semicircle at infinity in the lower half plane
    describing retarded propagation in time.

    In order to understand the nature of the singularities of the
    propagator, we must first address the structure of the self
    energy, in particular the imaginary part, which determines the
    relaxation rates. Again we focus on negative helicity neutrinos
    for simplicity. Upon the analytic continuation
    $s=-i\omega+\epsilon$   for this case we define

    \be S(k,\omega) \equiv S_{h=-1}(k;s=-i\omega+\epsilon) =
    (\omega+k)\, \frac{1}{4} \left.\mathrm{Tr}(\gamma^0-\vec{\gamma}\cdot
    \hat{ \textbf{ k}})
    \widetilde{\Sigma}_{aa}(\vk,s)\right|_{s=-i\omega+\epsilon} \ee

 From equation (\ref{disc}) which is a consequence of the dispersive
 form (\ref{siglapla}) of the self energy $\widetilde{\Sigma}_{aa}(\vk,s)$ ,
 it  follows that \be  S(k,\omega)=   S_R(k,\omega) + i
 S_I(k,\omega) \label{reimS}\ee where $S_{R,I}$ are the real and
 imaginary parts respectively. The real part of the self energy
 determines the correction to the dispersion relations of the
  neutrino \emph{quasiparticle} modes in the medium, namely the ``index of
 refraction'', while the imaginary part determines the relaxation
 rate of these quasiparticles.

    \subsection{ The self-energy: quasiparticle dispersion relations and widths:}
\begin{figure}[ht!]
\begin{center}
\epsfig{file=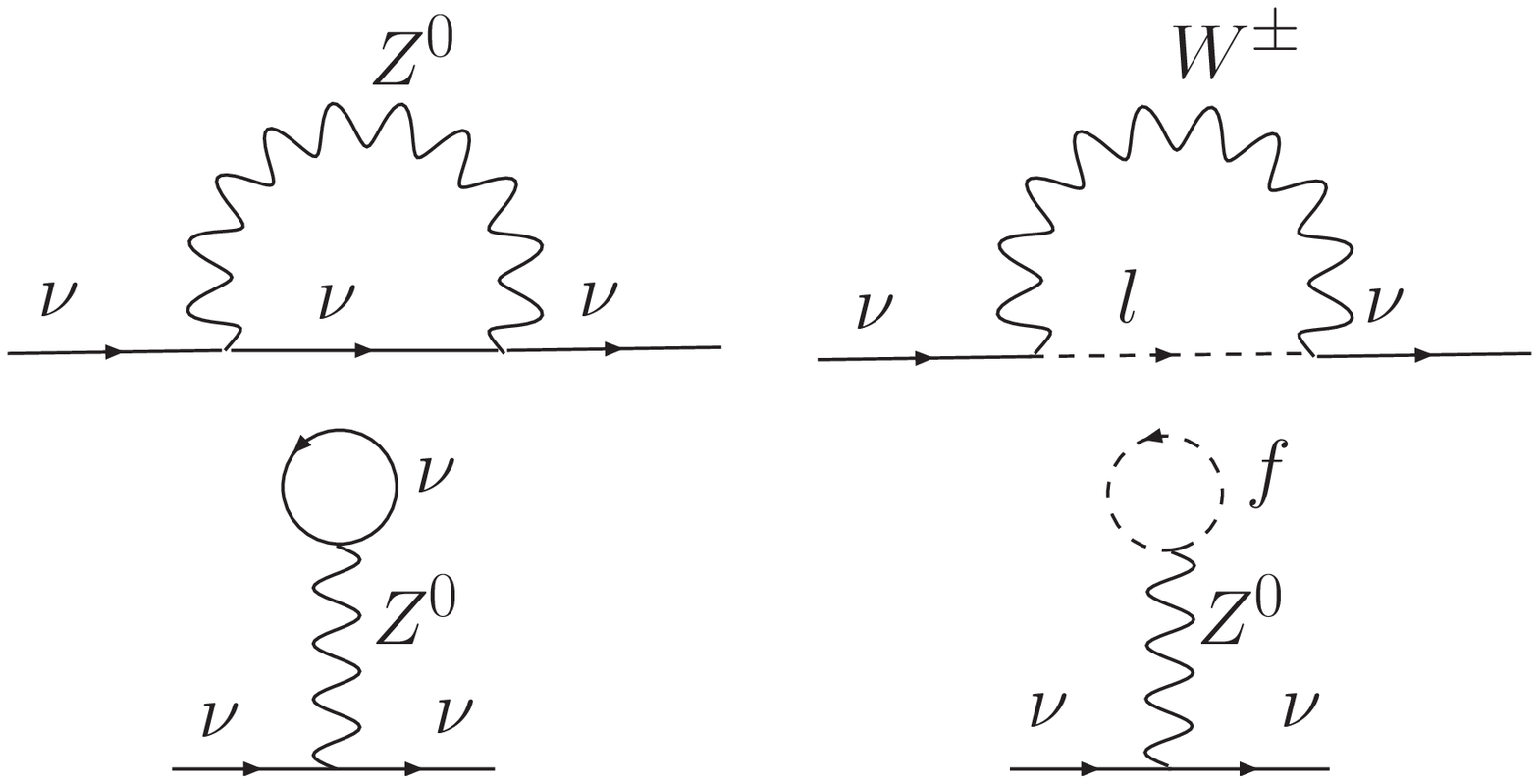,height=3.5in,width=3.5in,keepaspectratio}
\epsfig{file=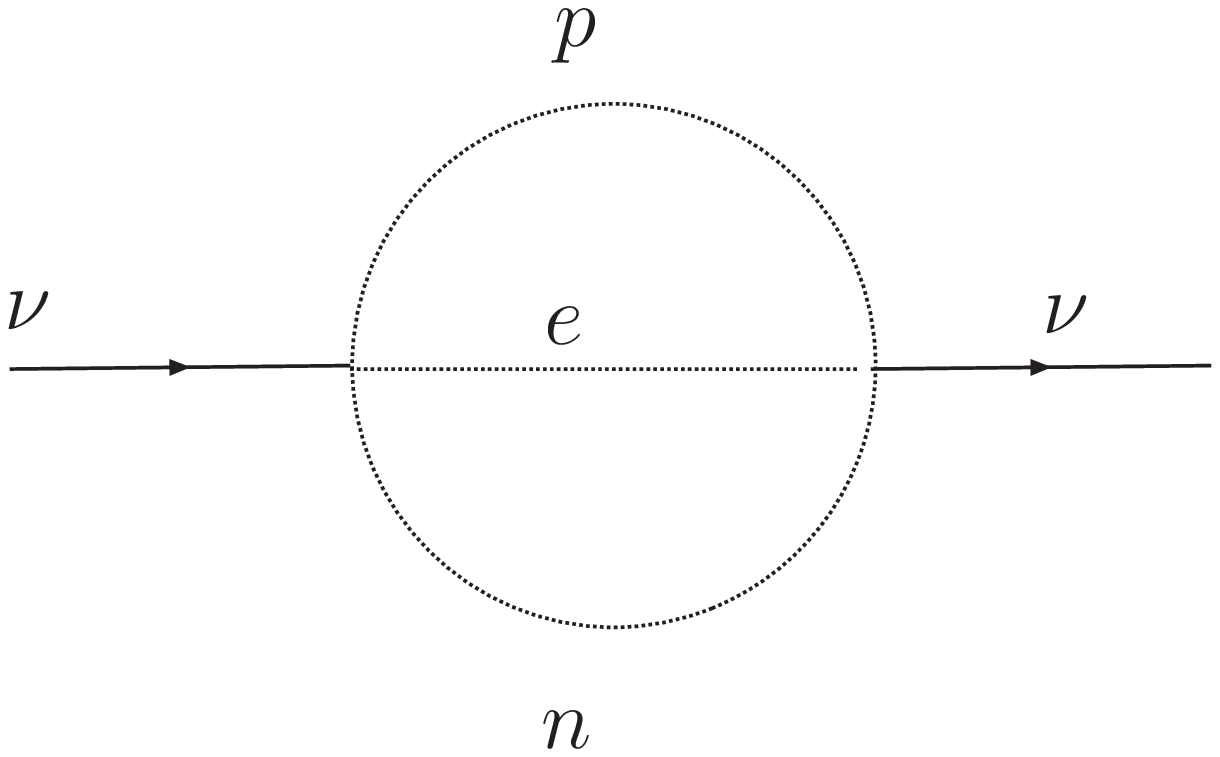,height=3.5in,width=2in,keepaspectratio}
 \caption{Left: one loop contributions to the self energy, the diagrams in the second line yield $\Sigma^{tad}$. These contributions
 are of $\mathcal{O}(G_F)$. Right: two
 loops contribution  of $\mathcal{O}(G^2_F)$ to the self-energy in Fermi's effective field theory limit, with internal lines corresponding to hadrons and the charged lepton,
 or alternatively quarks and the charged lepton above the QCD phase transition.  }
  \label{fig:selfenergy}
\end{center}
\end{figure}

      Figure (\ref{fig:selfenergy}) shows the one loop contributions of $\mathcal{O}(G_F)$ including
    the neutral current tadpole diagrams which contribute to the in-medium ``index of refraction'' for one active
    species, and the  two loop contribution of $\mathcal{O}(G^2_F)$ with intermediate states of hadrons (or quarks)
     and the associated  charged lepton, in the limit of Fermi's effective field theory.

     In a medium at temperature $T$ the real part of the one-loop contributions to $S(k,\omega)$ is  of the
     form\cite{notzold,eunu,dolgovrev,kev1}

     \be  S_R(k,\omega) =  (\omega+k)\,  G_F T^3 \left[L+
     \frac{T}{M^2_W}\left(a \omega + b k \right) \right] \label{ReS}
     \ee where $L$ is a   function of the asymmetries of the
     fermionic species and $a,b$ simple coefficients, all of which
     may be read from the results in ref.\cite{notzold,kev1,eunu}.
     \emph{Assuming} that all asymmetries are of the same order as
     the baryon asymmetry in the early Universe $L \sim 10^{-9}$ the
     term $\propto T/M^2_W$ in (\ref{ReS}) for $\omega \sim k \sim
     T$ dominates over the asymmetry term for $T \gtrsim
     3\,\mathrm{MeV}$\cite{notzold,eunu} and in what follows we neglect the CP violating terms associated with
     the lepton asymmetry.

\begin{figure}[ht!]
\begin{center}
\epsfig{file=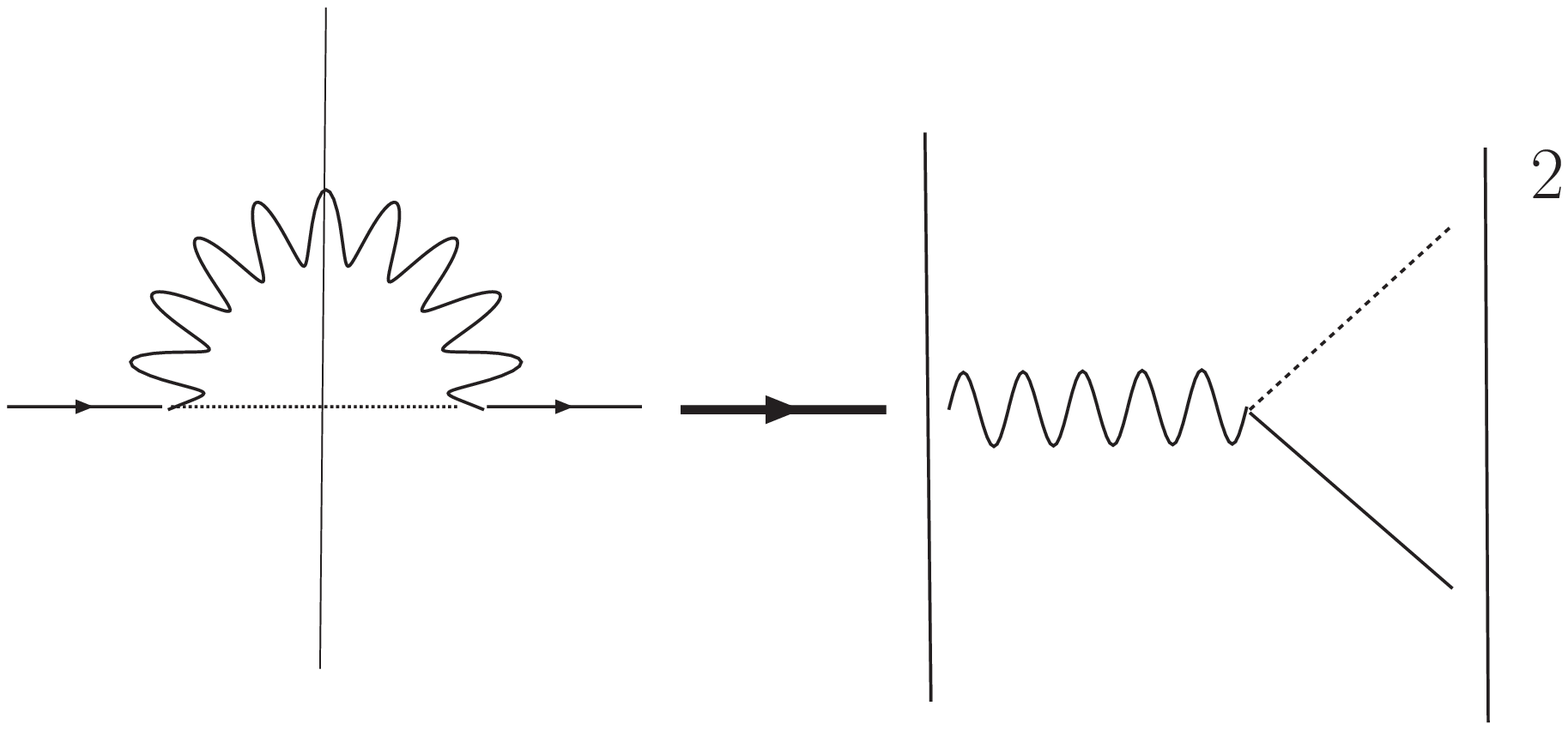,height=3.5in,width=3.5in,keepaspectratio}
\epsfig{file=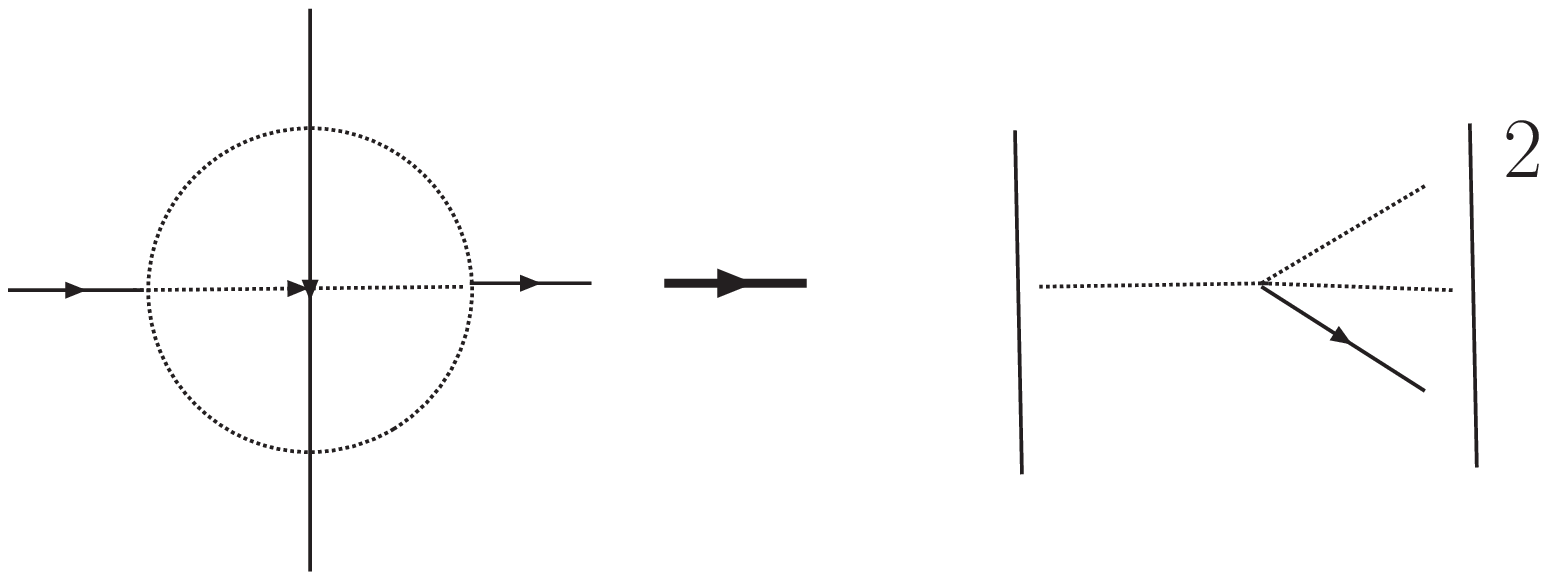,height=3.5in,width=3.5in,keepaspectratio}
 \caption{Contributions to the imaginary part of the self energy, the vertical line represents a Cutkosky cut. Left: discontinuity from the one loop contributions to the self energy
  of $\mathcal{O}(G_F)$,   from the decay of vector bosons for example $W \rightarrow l \overline{\nu}$. Right: discontinuity from the two
 loops contribution  of $\mathcal{O}(G^2_F)$, arising for example from $n\rightarrow p+e^+ +\overline{\nu}_e$ or similar processes at the
 quark level.  }
  \label{fig:selfenergydisc}
\end{center}
\end{figure}

      The
     imaginary part to one loop order is obtained from a Cutkosky
     cut (discontinuity) of the diagrams with vector boson exchange
     shown on the left side in figure (\ref{fig:selfenergydisc}) and is determined by
     the processes $W \rightarrow l_a \overline{\nu}_a, Z
     \rightarrow \overline{\nu}_a \nu_a$. Both of these contributions
     are exponentially suppressed at temperatures $T\ll M_{W,Z}$,
     hence the one-loop contributions to the imaginary part of
     $S(k;\omega)$ is \emph{negligible} for temperatures well below the
     electroweak scale. The two loop contribution to the imaginary
     part is obtained from the discontinuity cut of the two loop
     diagram with internal hadron or quark and charged lepton lines in
     figure (\ref{fig:selfenergy}). Some of the processes that contribute to
     the imaginary part in this order are for example neutron $\beta$ decay
     $n\rightarrow p+e^+ + \overline{\nu}$ and its inverse, along
     with scattering processes in the medium. The imaginary part of
     the   self-energy for these contributions on-shell is
     proportional to $G^2_F \,k \, T^4$\cite{notzold,dolgovrev} at temperatures $T \ll M_W$.
     Therefore in this temperature range

     \be      S_I(k,\omega) \sim   (\omega+k)\,  G^2_F \, k \, T^4 \,.  \label{ImS}
     \ee The consistency and validity of perturbation theory and of Fermi's
     effective field theory for scales $\omega, k, T \ll M_W$ entail
     the following inequality

     \be  S_I(k,\omega) \ll  S_R(k,\omega)\,.
     \label{ineq}\ee For example near the neutrino mass shell for
     ultrarelativistic neutrinos with $\omega \sim k$, assuming $L \sim 10^{-9}$ and discarding this
     CP violating contribution for $T > 3
     \,      \mathrm{MeV}$ because it is subleading, we find

     \be \frac{ S_I(k,\omega)}{ S_R(k,\omega)} \sim
     g^2 \label{pert}\ee with $g$ the standard model weak   coupling. This discussion
     is relevant for the detailed understanding when the quantum Zeno effect is operative (see section (\ref{sec:implications}) below).

  The propagator $G(k;\omega)$ for negative helicity neutrinos is
     found to be given by

     \be \label{Gpole} G(k;\omega) = \frac{1}{2\beta } \Bigg[\frac{1}{\alpha-\beta}-\frac{1}{\alpha+\beta}
     \Bigg] \,  \left(
          \begin{array}{cc}
          \alpha+\beta \cos 2\theta_m &  - \beta\sin 2\theta_m   \\
             - \beta \sin 2\theta_m &  \alpha - \beta \cos 2\theta_m  \\
          \end{array} \right)\,,\ee where we have suppressed the
          arguments for economy of notation, and defined

          \bea \alpha & = & \omega^2 - k^2
          -\overline{M}^{\,2}+\frac{1}{2} S_R(k,\omega) +
          \frac{i}{2} S_I(k,\omega) \,,\label{alfa2} \\
          \beta & = & \frac{\delta M^2}{2} \Bigg[\Big(\cos2\theta -
          \frac{S_R(k,\omega)}{\delta M^2}-i\frac{S_I(k,\omega)}{\delta
          M^2}\Big)^2+\sin^22\theta\Bigg]^{\frac{1}{2}}\,,
          \label{beta2}\\ \theta_m & \equiv &  \theta^{(-1)}_m(k,s=-i\omega+\epsilon)\,.\label{tetam}\eea
The inequality (\ref{ineq}) licenses us to write $\beta$
consistently up to $\mathcal{O}(G^2_F)$ as

\be \label{betaapx} \beta \simeq \frac{\delta M^2}{2}\rho(k,\omega)
-\frac{i}{2}S_I(k,\omega) \cos2\theta_m \,,\ee where \be
\rho(k,\omega) = \Bigg[\Big(\cos2\theta -
          \frac{S_R(k,\omega)}{\delta M^2}
          \Big)^2+\sin^22\theta\Bigg]^{\frac{1}{2}}\,. \ee

Equation (\ref{Gpole}) makes manifest that $G(k;\omega)$ is strongly
peaked at the values of $\omega$ for which $\alpha = \pm \beta$.
These determine the position of the complex poles in the analytic
continuation. In the relativistic approximation $k \gg M_{1,2}$ we
find:

\begin{itemize}
\item{ For {\bf $\alpha = \beta$: } \be \omega_1(k) = E_1(k) - i \, \frac{\Gamma_1(k)}{2}
\label{omegap1} \ee
 with \bea E_1(k) & \approx &  k +  \frac{1}{2k}
\left[ \overline{M}^{\,2} + \frac{\delta M^2}{2}
\rho(k)-\frac{S_R(k)}{2}\right]\label{E1}\\\frac{\Gamma_1(k)}{2}  &
= & \frac{\Gamma_{aa}(k)}{2} \cos^2\theta_m(k)\,  \label{gamma1}\eea
}

\item{ For {\bf $\alpha = -\beta$: } \be \omega_2(k) = E_2(k) - i
\frac{\Gamma_2(k)}{2}\label{omegap2}  \ee with \bea E_2(k) & \approx
& k +  \frac{1}{2k} \left[ \overline{M}^{\,2} - \frac{\delta M^2}{2}
\rho(k)-\frac{S_R(k)}{2}\right]\label{E2}\\ \frac{\Gamma_2(k)}{2} &
=  & \frac{\Gamma_{aa}(k)}{2} \sin^2\theta_m(k)\,.\label{gamma2}\eea
}
\end{itemize}

\noindent where \be \rho(k) \equiv \Bigg[\Big(\cos2\theta -
          \frac{S_R(k,\omega=k)}{\delta M^2}
          \Big)^2+\sin^22\theta\Bigg]^{\frac{1}{2}} \,, \label{rhoofk}\ee  and  $\Gamma_{aa}(k)$
is the
 standard model result for the scattering rate of the active neutrino species\cite{kev1,notzold,foot,dibari}
\be \frac{\Gamma_{aa}(k)}{2}= \frac{S_I(k,\omega=k)}{2k} =
\frac{1}{4} \mathrm{Tr}(\gamma^0-\vec{\gamma}\cdot
    \hat{ \textbf{ k}})
    \mathrm{Im}{\Sigma}_{aa}(\vk,\omega=k) \sim G^2_F \,T^4\,k
\label{gammaa}\ee   and $\theta_m(k) = \theta^{(h=-1)}_m(k,s=-ik)$
is the mixing angle in the medium for negative helicity   neutrinos
of energy $\omega \sim k$ in the relativistic limit. The relations
(\ref{gamma1},\ref{gamma2}) are the same as those recently found in
reference\cite{recentours}.

Combining all the results    we find

\bea
     \left(
       \begin{array}{c}
         \nu_a(\vk,t) \\
         \nu_s(\vk,t)  \\
       \end{array}
     \right)& & =  \Bigg[ e^{-iE_1(k)t}\,e^{-\frac{\Gamma_1(k)}{2} t}\,  \frac{1}{2} \left(
                                                                            \begin{array}{cc}
                                                                              1+\cos 2\theta_m(k) & -\sin2\theta_m(k) \\
                                                                              -\sin2\theta_m(k) & 1-\cos 2\theta_m(k) \\
                                                                            \end{array}
                                                                          \right)+ \nonumber \\ &&  e^{-iE_2(k)t}\,e^{-\frac{\Gamma_2(k)}{2}\,t} \,
                                                                          \frac{1}{2} \left(
                                                                            \begin{array}{cc}
                                                                              1-\cos 2\theta_m(k) &  \sin2\theta_m(k) \\
                                                                               \sin2\theta_m(k) & 1+\cos 2\theta_m(k) \\
                                                                            \end{array}
                                                                          \right)
        \Bigg] \left(
       \begin{array}{c}
         \nu_a(\vk,0) \\
        \nu_s(\vk,0)  \\
       \end{array}
     \right) \label{expoft} \eea

     This expression can be written in the following  more illuminating
     manner,

     \be
     \left(
       \begin{array}{c}
         \nu_a(\vk,t) \\
         \nu_s(\vk,t)  \\
       \end{array}
     \right) =   U(\theta_m(k))   \left(
                                                                            \begin{array}{cc}
                                                                              e^{-iE_1(k)t}\,e^{-\frac{\Gamma_1(k)}{2} t} & 0 \\
                                                                              0 & e^{-iE_2(k)t}\,e^{-\frac{\Gamma_2(k)}{2}\,t} \\
                                                                            \end{array}
                                                                          \right)   U^{-1}(\theta_m(k))      \left(
       \begin{array}{c}
         \nu_a(\vk,0) \\
        \nu_s(\vk,0)  \\
       \end{array}
     \right)\,, \label{expoft2} \ee where $U(\theta_m(k))$ is the
     mixing matrix (\ref{vacrot}) but in terms of the mixing angle
     in the medium.

     In obtaining the above expressions we have neglected
     perturbative corrections from wave function renormalization and
     replaced $\omega + k \sim 2 k$ thus   neglecting terms that
     are subleading in the relativistic limit, and the imaginary part in $\omega$, which
     although it is of $\mathcal{O}(G^2_F)$,    yields the effective Wigner-Weisskopf approximation\cite{recentours}.

     \subsection{  Physical interpretation:}\label{subsec:phys} The above results have the following clear
     physical interpretation. The active (a) and sterile (s)
     neutrino fields in the medium are linear combinations of the
     fields associated with the $1,2$ quasiparticle modes with dispersion relations $E_{1,2}(k)$
     and damping rates $\Gamma_{1,2}(k)$
      respectively, on the mass shell of the quasiparticle modes the
      relation between them is the usual one for neutrinos
      propagating in a medium with an index of refraction, namely
     \bea
     \nu_a(\vec{k},t) & = &  \cos\theta_m(k)~ \nu_1(\vec{k},t) +
     \sin\theta_m(k)~
     \nu_2(\vec{k},t) \\\nu_s(\vec{k},t) & = &  \cos\theta_m(k) ~\nu_2(\vec{k},t) -
     \sin\theta_m(k)~
     \nu_1(\vec{k},t)\,. \label{quasicombo}\eea

 These relations between the expectation values of flavor fields and
 the fields associated with the propagating quasiparticle modes in
 the medium are obtained from the diagonalization of the neutrino
 propagator on the mass shell of the quasiparticle modes. These are
 recognized as the usual relations between flavor and ``mass''
 fields in a medium with an index of refraction.

     At temperatures much
     higher than that at which a resonance occurs (and for $k \sim
     T$) $\theta_m(k)\sim \pi/2$ then   $\nu_a \sim \nu_2$, and
     the active neutrino features a damping rate $\Gamma_2 \sim
     \Gamma_{aa}$ while the sterile neutrino $\nu_s \sim \nu_1 $
     with a damping rate $\Gamma_1 = \Gamma_{aa}\cos^2\theta_m(k)
     \ll \Gamma_{aa}$. In the opposite limit for temperatures much
     lower than that of the resonance and for very small vacuum
     mixing angles $\nu_a \sim \nu_1$ and features a damping rate
     $\Gamma_1 \sim \Gamma_{aa}$ while $\nu_s \sim \nu_2$ with a
     damping rate $\Gamma_2 \sim  \Gamma_{aa}\sin^2\theta \ll
     \Gamma_{aa}$. Thus it is clear that in both limits the active
     neutrino has the larger damping rate and the sterile one the
     smallest one. This physical interpretation confirms that there
     \emph{must} be two widely different time scales for relaxation
     in the  high and low temperature limits, the \emph{longest time scale} or alternatively the
     \emph{smallest damping rate} always corresponds to the sterile
     neutrino. This is obviously in agreement with the expectation
     that sterile neutrinos are much more weakly coupled to the plasma than the
     active neutrinos for $\sin2\theta_m(k) \sim 0$. This analysis
     highlights that \emph{two time scales} must be expected on
     physical grounds, not just one, the decoherence time scale,
     which only determines the suppression of the \emph{overlap}
     between the propagating states in the mixed neutrino state.

     The solution (\ref{expoft}) can  also be obtained from an effective
     Schroedinger-like equation but with
     an effective Weisskopf-Wigner Hamiltonian\cite{recentours}, not for the quantum states,
      which are \emph{not} decaying, but
      for the \emph{ensemble averages},

     \be i \frac{d}{dt} \left(
       \begin{array}{c}
         \nu_a(\vk,t) \\
         \nu_s(\vk,t)  \\
       \end{array}
     \right) = \mathcal{H}_{ww} \left(
       \begin{array}{c}
         \nu_a(\vk,t) \\
         \nu_s(\vk,t)  \\
       \end{array}
     \right)\label{scheq} \ee where the Weisskopf-Wigner effective
     Hamiltonian is \be \mathcal{H}_{ww} =
     \Big(k+\frac{\overline{M}^2}{2k}\Big)\mathbb{I}+ \frac{\delta M^2}{4k}
     \left(
       \begin{array}{cc}
          \cos2\theta & -\sin2\theta \\
         -\sin2\theta & -\cos2\theta \\
       \end{array}
     \right) - \frac{1}{2k}\Big(S_R(k) + i S_I(k) \Big) \left(
                                                          \begin{array}{cc}
                                                            1 & 0 \\
                                                           0 & 0 \\
                                                          \end{array}
                                                        \right)\,.
                                                        \label{Hww}
                                                        \ee
Although this form of the time evolution looks similar to the usual
quantum mechanical one, we emphasize that this equation of motion is
for the ensemble averages of the neutrino operators and $S_{R,I}$
are obtained from the self-energy on the mass shell of
ultrarelativistic neutrinos. This self energy  is an \emph{ensemble
average} of correlation functions of  the operators that describe
the degrees of freedom in the medium. This time evolution is an
\emph{a posteriori} consequence of the field-theoretical analysis,
which unambiguously reveals the time evolution. While the
description afforded by the Schroedinger-like equation for the
ensemble averages with a \emph{non-hermitian} Weisskopf-Wigner
Hamiltonian is simpler, it can only be rigorously justified through
the detailed study presented above.

     The emergence of \emph{two
     time scales} can also be gleaned in the pioneering work on sterile neutrino production of
     ref.\cite{cline} (see eqn. (10) in that reference) which in this reference where obtained
     within a phenomenological Wigner-Weisskopf approximation akin to (\ref{scheq},\ref{Hww}). Our
     quantum field theory study based on the full density matrix and the
      neutrino propagator in the medium provides a consistent and systematic
     treatment of propagation in the medium that displays both time
     scales and provides a derivation of the effective Schroedinger-like evolution with the Weisskopf-Wigner
     Hamiltonian\cite{recentours}.

\section{Quantum Zeno effect}\label{sec:QZE}
\subsection{Real time interpretation and general conditions}\label{subsec:QZrt}

Consider  a density matrix in which the expectation value of the
active neutrino field is non-vanishing, but that of the sterile
neutrino field \emph{vanishes} at the initial
     time $t=0$, namely $\nu_a(\vk,0)\neq 0~;~\nu_s(\vk,0)=0$. Then  it is clear from equation (\ref{expoft}) that
     flavor off-diagonal density matrix elements develop in time
     signaling that
     sterile neutrinos are produced \emph{via active-sterile mixing}
     with amplitude

     \be \nu_s(\vk,t) = -\frac{1}{2} \sin2\theta_m(k) \Bigg[ e^{-iE_1(k)t}\,e^{-\frac{\Gamma_1(k)}{2}
     t}\,-  e^{-iE_2(k)t}\,e^{-\frac{\Gamma_2(k)}{2}\,t} \Bigg]
     \nu_a(\vk,0) \label{nusprod} \ee  From the solution (\ref{nusprod}) we
     introduce the
     \emph{generalized  transition probability in the medium} from
     the expectation values of the neutrino fields in the density
     matrix, these are the transition probabilities between ensemble
     averages of one-particle states of the neutrino fields,

     \be P_{a\rightarrow s}(t) =
     \Bigg|\frac{\nu_s(\vk,t)}{\nu_a(\vk,0)}\Bigg|^2=
     \frac{\sin^22\theta_m(k)}{2}\,e^{-\Gamma(k)t}\,
     \Bigg[\cosh(\gamma(k) t)- \cos(\Delta E(k) t) \Bigg]
     \label{proboft} \ee where

     \bea \Gamma(k) & = & \frac{1}{2}\Big(\Gamma_1(k)+\Gamma_2(k)\Big)= \frac{\Gamma_{aa}(k)}{2}
     \label{gammatotal} \\
      \gamma(k) & = & \frac{1}{2}\Big(\Gamma_1(k)-\Gamma_2(k)\Big) = \frac{\Gamma_{aa}(k)}{2} \cos 2\theta_m(k)
       \label{gammafin} \\
     \Delta E(k) & = & E_1(k)-E_2(k)=\frac{\delta
     M^2}{2\,k} \rho(k) \label{deltaomeg} \eea For the analysis that
     follows it is more convenient to write (\ref{proboft}) in the
     form \be P_{a\rightarrow s}(k;t) =
     \frac{\sin^22\theta_m(k)}{4} \,
     \Bigg[ {e^{-\Gamma_1(k)\,t}} + {e^{-\Gamma_2(k)\,t}} \,-  2\, e^{-\Gamma(k)t}\cos(\Delta E(k) t)
     \Bigg]\,.
     \label{proboft2} \ee The first two terms are obviously the
     probabilities for the quasiparticle modes $1,2$, while the
     oscillatory term is the usual interference between these but
     now damped by the factor $e^{-\Gamma(k)t}$. This form of the
     transition probability is remarkably similar to the familiar
     transition probability in $K_0-\overline{K}_0$ or $B_0-\overline{B}_0$     oscillations\cite{kaons,raiden}.
     However we emphasize that (\ref{proboft2}) is the
     \emph{generalized} transition probability extracted from
     expectation values in the density matrix.

     We highlight that
     the \emph{decoherence} time scale is precisely
     $\Gamma^{-1}(k)=2/\Gamma_{aa}(k)$ as anticipated in
     references\cite{stodolsky,foot}, since the \emph{interference}
     between the two quasiparticle modes is suppressed on this time
     scale. However, the \emph{total} transition probability is
     suppressed on this time scale \emph{only} if $\Gamma_1=
     \Gamma_2 =\Gamma$, namely near a resonance. In this case
\be P_{a\rightarrow s}(t) =
      \sin^22\theta_m  ~ e^{-\frac{\Gamma_{aa}}{2}\,t}\,\sin^2\left[\frac{\Delta
      E }{2}\,
      t\right]
     \,,
     \label{proboftres} \ee which is the result quoted in
     reference\cite{foot} (see eqn. (\ref{Pasfoot})). Under these conditions quantum Zeno
     suppression occurs when $\Gamma(k) \gg \Delta E(k)$ in
     which case the decoherence time scale is much smaller than the
     oscillation time scale and the transition probability is
     suppressed before $a\rightarrow s$ oscillations take place.

\begin{figure}[ht!]
\begin{center}
\epsfig{file=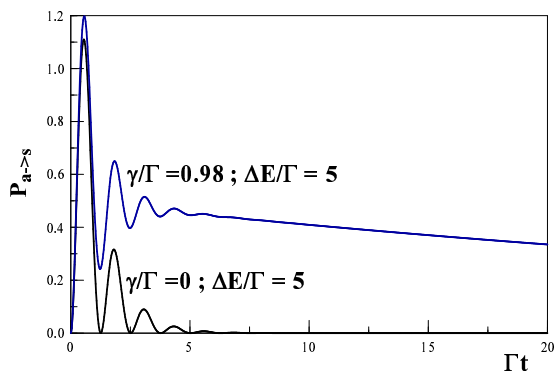,height=3in,width=3in,keepaspectratio}
 \caption{The transition probability $P_{a\rightarrow s}(t)$
 (without the prefactor $\sin^22\theta_m/2$) vs. $\Gamma t$. The figure
  depicts the cases $\cos2\theta_m(k) = 0.98$
 and $\cos2\theta_m(k) = 0 $ respectively, both with
 $\Delta E/\Gamma = \delta M^2\rho(k)/2k\Gamma =5$.
  The scale for suppression of the oscillatory interference is
 $1/\Gamma$ in both cases. }
  \label{fig:pas5}
\end{center}
\end{figure}

\begin{figure}[ht!]
\begin{center}
\epsfig{file=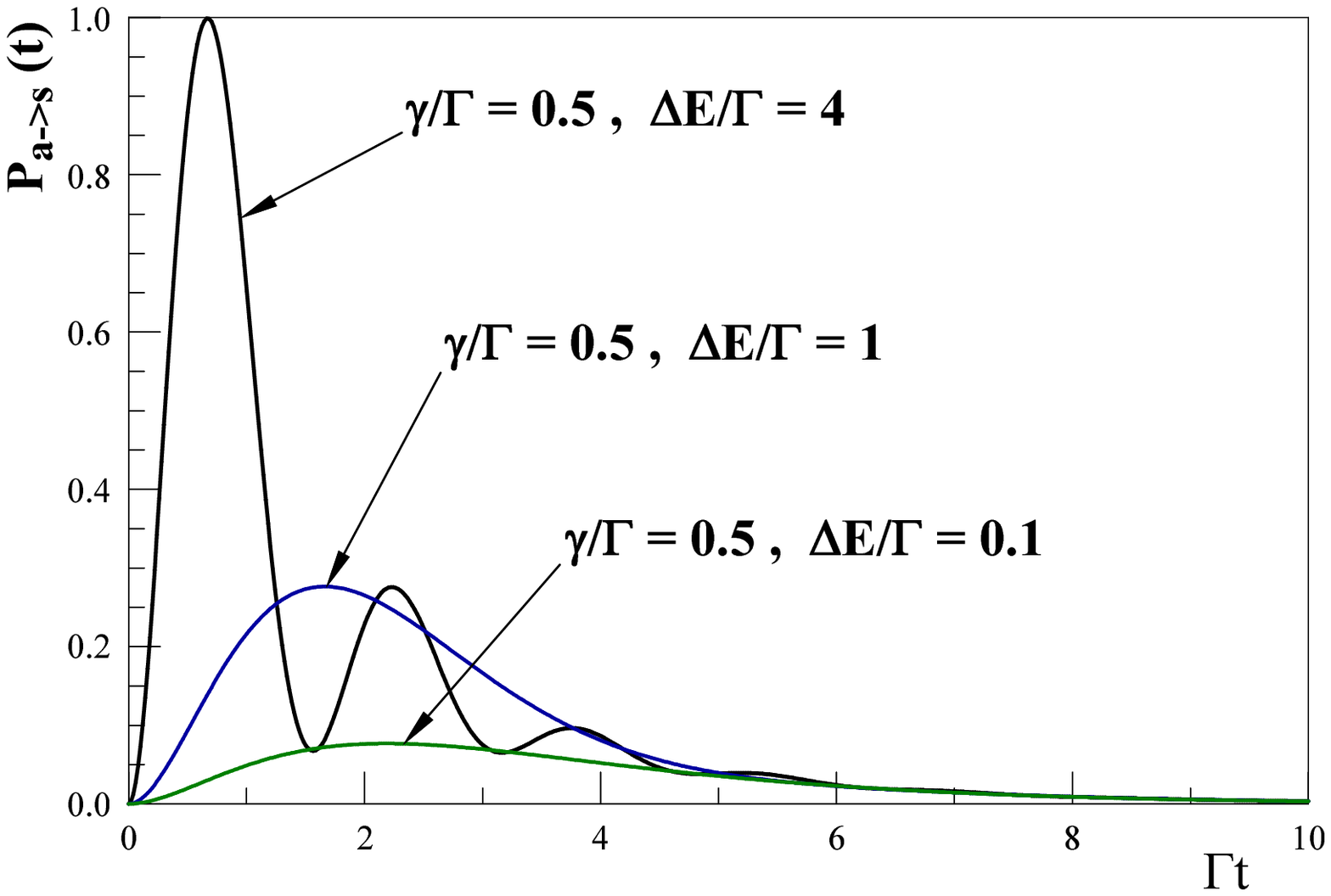,height=3in,width=3in,keepaspectratio}
\epsfig{file=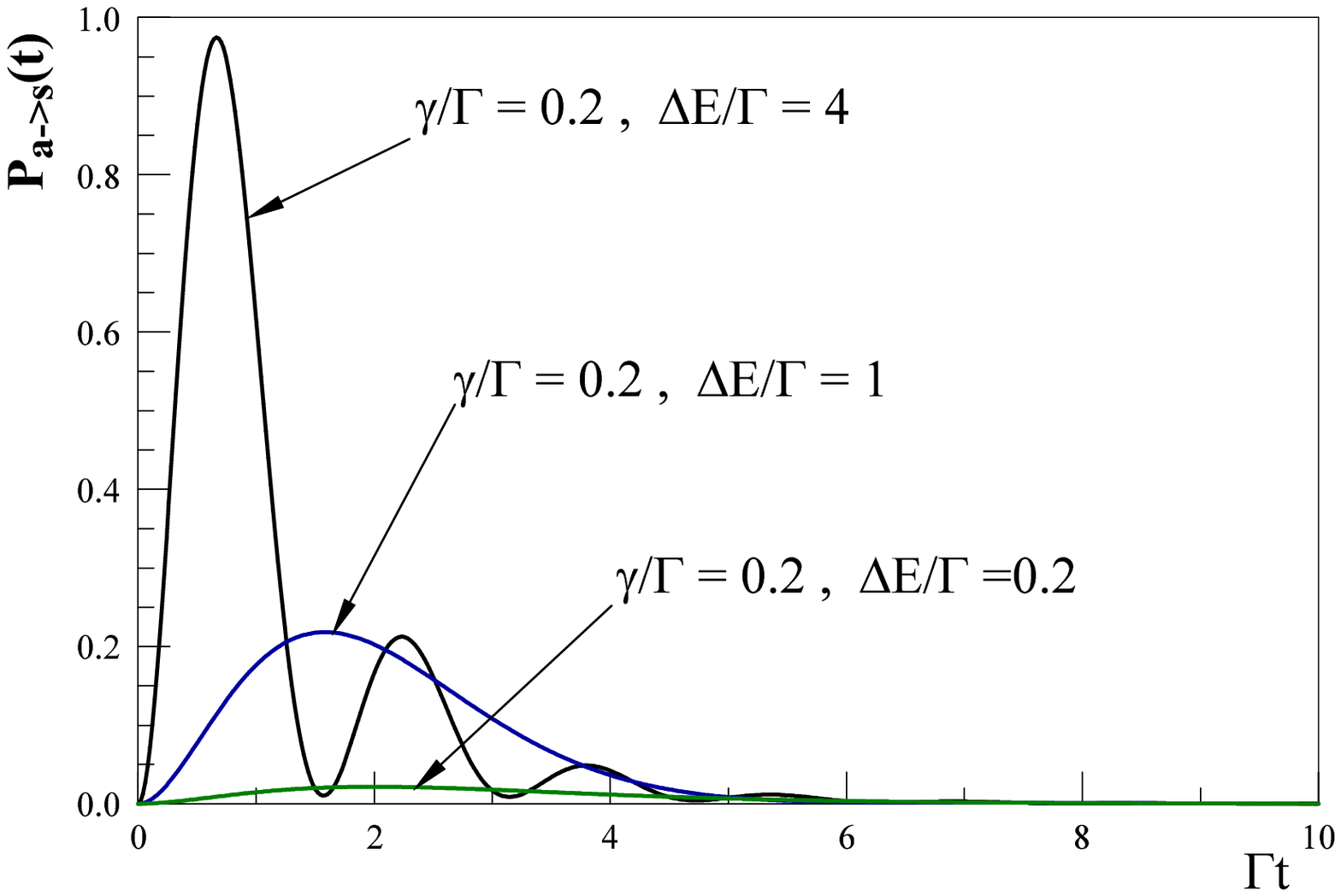,height=3in,width=3in,keepaspectratio}
 \caption{The transition probability $P_{a\rightarrow s}(t)$   (without the prefactor $\sin^22\theta_m/2$) vs. $\Gamma t$.
 The left panel is for $\gamma/\Gamma=0.5$, $\Delta E/\Gamma=4,1,0.1$, the right panel is for $\gamma/\Gamma=0.2$, $\Delta E/\Gamma=4,1,0.2$. }
  \label{fig:patosoft}
\end{center}
\end{figure}

    However, far away on
     either side of the resonance, although the oscillatory interference term is suppressed
     on the decoherence time scale $\Gamma^{-1}$,  the transition probability is
     \emph{not} suppressed on this   scale but on a
     much longer time scale, determined by the \emph{smaller}
     of $\Gamma_{1,2}$. Only when $\Gamma_1=\Gamma_2 =\Gamma$,
     namely $\gamma=0$ both the coherence (oscillatory interference
     term) and the transition probability are suppressed on the
     decoherence time scale. This phenomenon is displayed in figures
     (\ref{fig:pas5},\ref{fig:patosoft} ) which show  the transition probability
     as a function of time without the  prefactor
     $\sin^22\theta_m(k)$ for several values of the ratios $\gamma/\Gamma;\Delta E/\Gamma$.

\begin{figure}[ht!]
\begin{center}
\epsfig{file=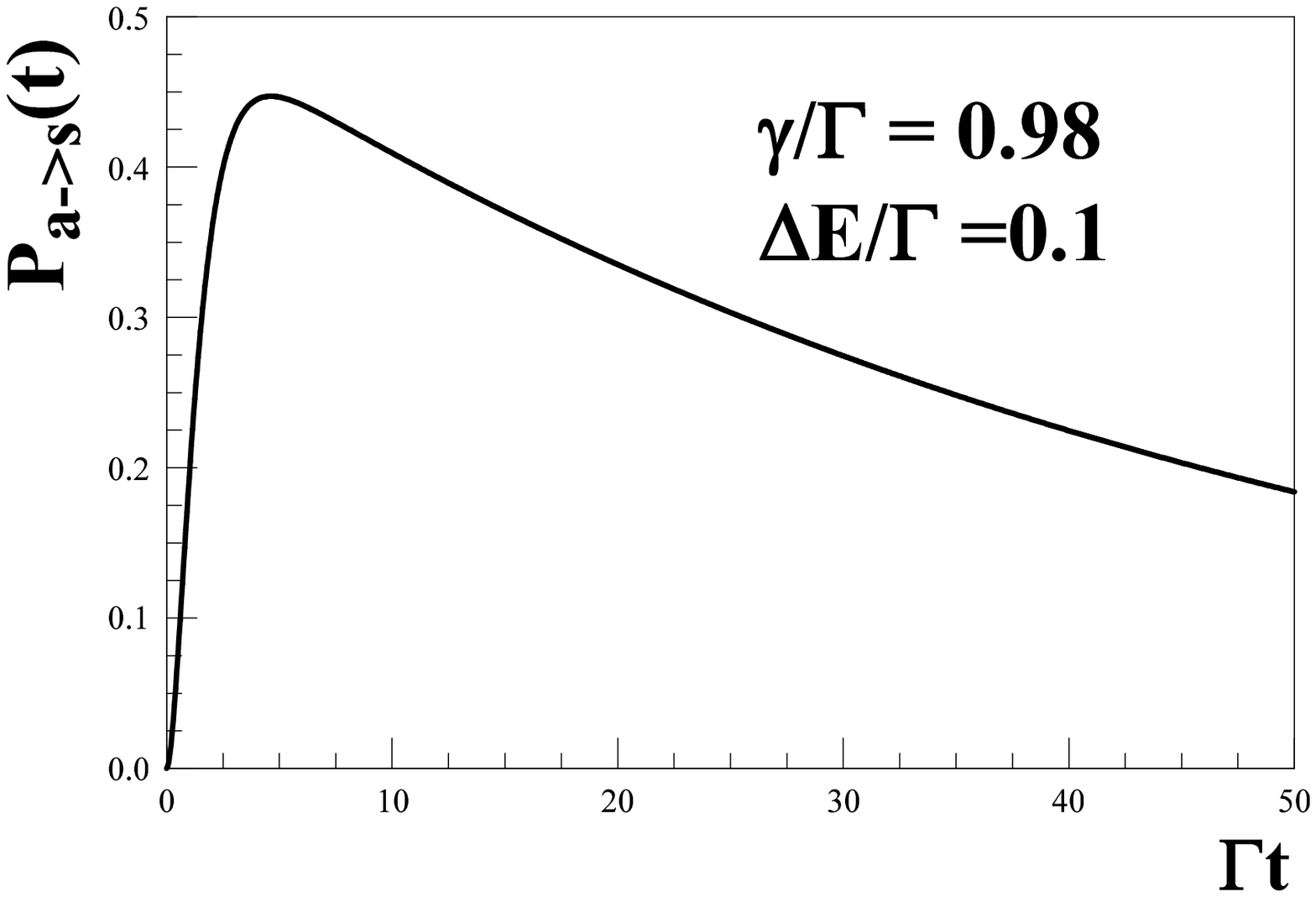,height=3in,width=3in,keepaspectratio}
\epsfig{file=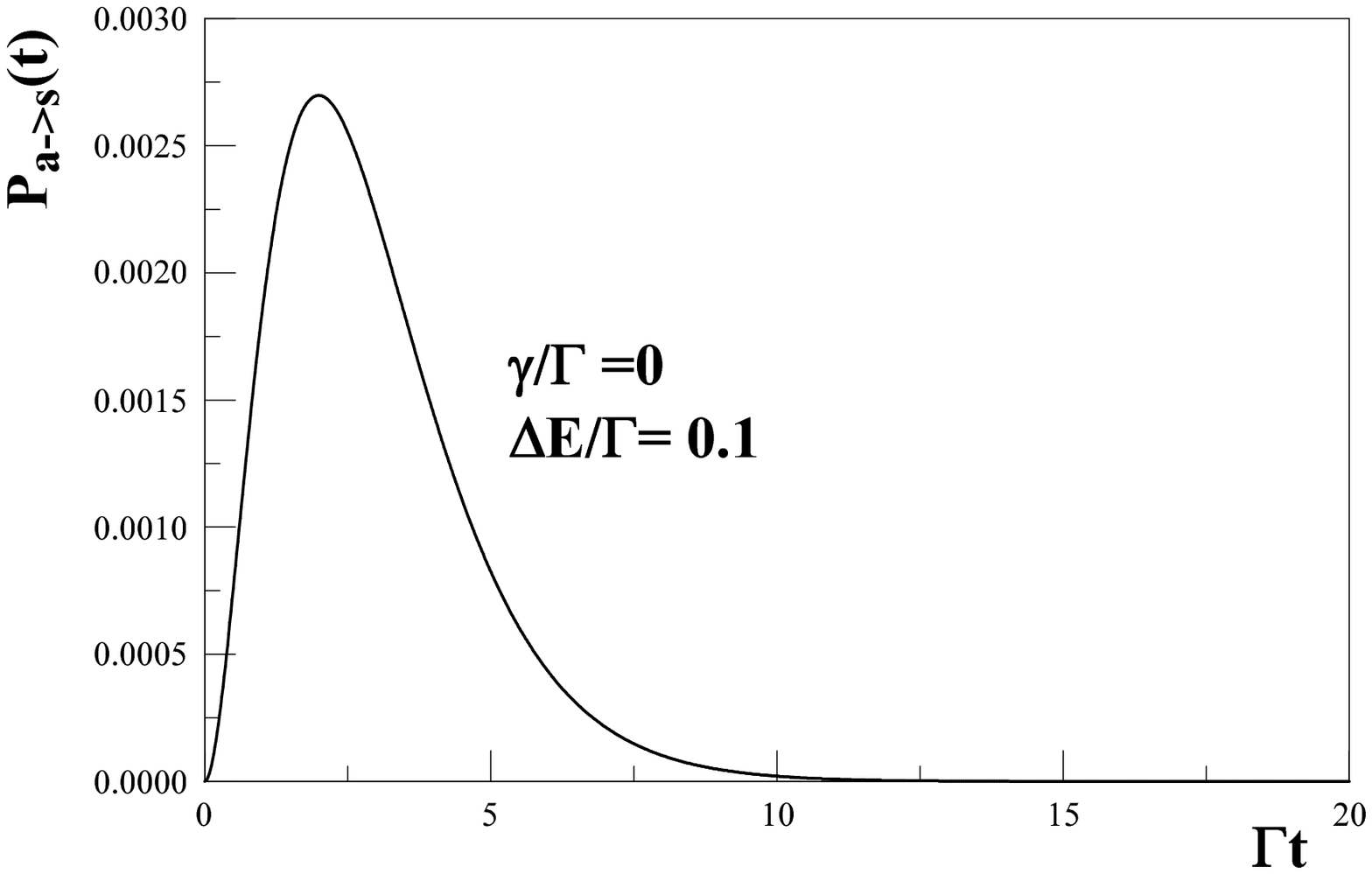,height=3.2in,width=3.2in,keepaspectratio}
 \caption{The transition probability $P_{a\rightarrow s}(t)$
  (without the prefactor $\sin^22\theta_m/2$) vs. $\Gamma t$ in the quantum Zeno limit $\Gamma \gg \Delta E$ for the cases $\cos2\theta_m(k) = 0.98,0$ and
 $\Delta E/\Gamma = \delta M^2\rho(k)/2k\Gamma =0.1$.
 The ramp-up time scale is $\sim 1/\Gamma_1\sim 1/\Gamma$. In the left figure the damping time
 scale is $\sim 1/\Gamma_2 \sim 50/\Gamma$. The right figure displays the resonant case for which the damping and coherence
 time scale coincide, when the conditions for quantum Zeno suppression are fulfilled. }
  \label{fig:pas}
\end{center}
\end{figure}

      Even for $\Gamma(k) \gg \Delta E(k)$, claimed in the literature \cite{stodolsky,kev1} to
      be the condition for  quantum Zeno suppression,
       the transition probability is substantial on time scales
     much longer than $\Gamma^{-1}$ if $\Gamma_1$ and $\Gamma_2$ are \emph{widely
     separated}, namely if $|\gamma/\Gamma|\sim 1$. This situation is depicted in figure
     (\ref{fig:pas}). From this analysis we conclude that the
     conditions for quantum Zeno suppression of $P_{a\rightarrow
     s}(t)$ are: {\bf i):} $\Gamma(k) \gg \Delta E(k)$ \emph{and}
     {\bf ii):)} $\gamma(k) \sim 0$, namely $\Gamma_1(k) \sim
     \Gamma_2(k)$. These conditions    are obtained directly from the time dependence of $P_{a\rightarrow s}(t)$
     without taking any time average.  We then emphasize that it is \emph{not}
     necessary to average the probability over time to recognize the criteria
     for the quantum Zeno effect, these can be directly gleaned from
     the time evolution of the probability as originally
     proposed\cite{misra}.

From the arguments in reference\cite{foot},   the effective sterile
neutrino production rate is obtained from the   average of the
transition probability on the decoherence time scale $\tau_{dec}$.
Using the result (\ref{proboft2}) we find instead

 \be \Big \langle P_{a\rightarrow s} \Big \rangle \equiv
 {\Gamma} \int^{\infty}_0 P_{a\rightarrow s}(t)\, dt  =  \frac{\sin^2
 2\theta_m}{2}~
  \frac{\left(\frac{\gamma}{ {\Gamma}}\right)^2 + \left(\frac{\Delta
E}{ {\Gamma}}\right)^2} {\left[1-\left(\frac{\gamma}{
{\Gamma}}\right)^2\right] \left[1+\left(\frac{\Delta E}{
{\Gamma}}\right)^2\right]}\,, \label{transaver}\ee where
$\Gamma,\gamma$ are given by eqns. (\ref{gammatotal},\ref{gammafin})
respectively.

This expression features two  remarkable differences with the result
(\ref{prob})\cite{foot}: the extra terms $(\gamma/\Gamma)^2$ in the
numerator and  $1-(\gamma/\Gamma)^2$ in the denominator, both are
consequence of the fact that the relaxation is determined by
\emph{two } time scales $\Gamma_1,\Gamma_2$. Only when these scales
are \emph{equal}, namely when $\gamma=0$ the result (\ref{prob})
often used in the literature is recovered.

We note that the result  (\ref{transaver}) is \emph{not} singular
since $1-(\gamma^2/\Gamma^2) = \sin^22\theta_m$, (see also the
discussion in section (\ref{subsec:timeav})).

This   analysis leads us to state that the \emph{complete
conditions} for quantum Zeno suppression  of the transition
probability are that \emph{both} $\gamma/\Gamma \ll 1$ \emph{and}
$\Delta E/\Gamma \ll 1$. That these are indeed the correct necessary
conditions for quantum Zeno suppression  can be gleaned from figures
(\ref{fig:patosoft}, \ref{fig:pas}) which display  the transition
probability (without the prefactor $\sin^22\theta_m/2$) as a
function of time for several values of the ratios
$\gamma/\Gamma,\Delta E/\Gamma$   without performing the time
average.

\subsection{High and low temperature limits: assessment of the quantum Zeno condition }

In order to establish when the quantum Zeno condition $\Delta
E(k)/\Gamma_{aa}(k) \ll 1$ is fulfilled we focus on the cases far
away from resonances and, according to the exhaustive analysis of
ref.\cite{dodelson,kev1} and the constraints from the X-ray
background in clusters\cite{boyarsky,hansen2},   in the region of
parameter space $1\,\mathrm{keV} \lesssim m_s \lesssim 10
\,\mathrm{keV}$ , $10^{-10} \lesssim \sin^22\theta \lesssim
10^{-6}$. We consider $T \gtrsim 3 \, \textrm{MeV}$ for which we can
neglect the CP violating asymmetry contribution in (\ref{ReS})
assuming that it is of the same order as the baryon asymmetry $L
\sim 10^{-9}$\cite{notzold,eunu}. In this regime $\delta M^2 \sim
m^2_s $, and from (\ref{ReS}) we find

\be \frac{S_R(k,k)}{\delta M^2} \sim 10^{-14} \,
\Bigg(\frac{T}{\textrm{MeV}} \Bigg)^6\, \Bigg(
\frac{k}{T}\Bigg)^2\,\Bigg( \frac{\textrm{keV}}{m_s}\Bigg)^2 \ee

Taking $k \sim T$ and $m_s \sim 1\,\textrm{keV}$ the MSW resonance
$S_R(k,k)/\delta M^2 =1$ occurs at $T_{MSW}\sim 215\,\textrm{MeV}$
(a more precise estimate yields $T \sim
180\,\textrm{MeV}$\cite{dodelson,kev1}). For $T \gg T_{MSW}$
corresponding to $ S_R(k,k)/\delta M^2 \gg 1$
  the active sterile oscillation frequency becomes

\be \Delta E(k) = \frac{\delta
     M^2}{2\,k} \rho(k) \sim \frac{S_R(k,k)}{2k} \sim \frac{G_F\,T^4\,k}{M^2_W} \ee

From the result (\ref{gammaa}) for $\Gamma_{aa}(k)$ we find in the
high temperature limit $T\gg T_{MSW}$

\be \frac{2\,\Delta E(k)}{\Gamma_{aa}(k)} \sim
\frac{G_F\,T^4\,k}{G^2_F\,T^4\,k\,M^2_W}\sim \frac{1}{g^2} \gg 1 \ee
where $g$ is the weak coupling. We note that in the high temperature
limit the ratio $\Delta E(k)/\Gamma_{aa}(k)$ becomes independent of
$T,k$. This result is in agreement with the conclusions in
ref.\cite{charged}.

In the low temperature limit $3\,\textrm{MeV} \lesssim T \ll
T_{MSW}$ it follows that $S_R(k,k)/\delta M^2 \ll 1$ and the
active-sterile oscillation frequency is \be \Delta E(k) \sim
\frac{m^2_s}{2k} \ee hence the ratio

\be \frac{\Delta E(k)}{\Gamma_{aa}(k)} \sim
\frac{m^2_s}{G^2_F\,T^4\,k^2} \sim 10^{16}
\,\Bigg(\frac{m_s}{\textrm{keV}}\Bigg)^2\,\Bigg(\frac{T}{\textrm{MeV}}
\Bigg)^{-6}\,
 \Bigg( \frac{k}{T}\Bigg)^{-2} \ee which for $k \sim T$ can be simplified to

 \be \frac{2\,\Delta E(k)}{\Gamma_{aa}(k)}\sim 10^{2} \Bigg(\frac{T_{MSW}}{T}\Bigg)^6 \gg 1\,.\ee

At the MSW resonance $T = T_{MSW}$, $\cos2\theta_m \sim 0$,  $\Delta
E(k) = m^2_s \sin2\theta/2k$ and  the ratio becomes

\be   \frac{2\,\Delta E(k)}{\Gamma_{aa}(k)}\sim 10^{2} \sin2\theta
\ll 1 \ee for $10^{-5} \lesssim \sin2\theta \lesssim 10^{-3}$.
Therefore at the MSW resonance $\cos2\theta_m(k)\sim 0$ and
\emph{both} conditions for quantum Zeno suppression, $\gamma/\Gamma
\ll1 $ , $\Delta E/\Gamma \ll 1$ are fulfilled. However, we point
out that near the resonance the oscillation frequency becomes
$\Delta E \sim \frac{\delta M^2}{2k} |\sin\theta|$ \emph{only} if
the second order corrections to the dispersion relations (real part
of the poles) are neglected, therefore \emph{if} $\Gamma \gg \Delta
E$ a reassessment of the perturbative expansion including second
order corrections to the dispersion relations \emph{is required},
since $\Gamma \propto G^2_F$.  Therefore this analysis leads us to
unambiguously conclude that with standard model interactions for the
active neutrino quantum Zeno suppression is \emph{not} realizable
either at high temperatures, when the matter potential dominates or
at very low temperatures where the mixing angle is close to the
vacuum value. Such possibility \emph{may only} emerge  very near an
MSW resonance, however for small mixing angle this case requires a
thorough reassessment of the dispersion relations in the medium
\emph{including} corrections of $\mathcal{O}(G^2_F)$ to the
oscillation frequency.

 \subsection{  Validity of the perturbative expansion:}

The quantum Zeno condition $\Gamma_{aa}(k) \gg \Delta E(k)$
 requires a consistent assessment of the validity of the
 perturbative expansion in the standard model and or Fermi's
 effective field theory. The active neutrino scattering rate
 $\Gamma_{aa} \propto G^2_F \,k \, T^4$ is a \emph{two loops} result, while to
 leading order in weak interactions,
 the index of refraction contribution to the dispersion relation
 $S_R(k,\omega)$ is of one-loop order\cite{notzold,eunu}.  In the high temperature   limit when
 $S_R \gg \delta M^2 \sim m^2_s$ the active-sterile oscillation frequency is

 \be \Delta E(k) \sim \frac{|S_R(k,k)|}{2k} \ee combining this result with equation
(\ref{gammaa}) at high temperature or density where the index of
refraction dominates over $\delta M^2$,  it follows that     \be
\frac{\Delta E(k)}{\Gamma_{aa}(k)} \sim
\frac{|S_R(k,k)|}{S_I(k,k)}\ee  for $k\sim T$ the perturbative
relation
 (\ref{pert}) states that this ratio is $ \gtrsim 1/g^2 \gg 1$ where $g$ is the weak gauge coupling. An opposite ratio, namely $\Delta E(k) / \Gamma_{aa} \ll 1$ would entail that
 the two-loop contribution ($\Gamma_{aa}$) is \emph{larger} than the one-loop contribution that yields the index of refraction
 $S_R$.   Thus quantum Zeno suppression at high temperature when the index of refraction dominates the oscillation frequency
 necessarily implies a breakdown of the  strict perturbative expansion.  Such potential breakdown of perturbation theory in the standard model or Fermi's effective
 field theory in the quantum Zeno limit has been already observed in a different context by these authors in ref.\cite{charged}, and deserves
  deeper scrutiny. We are currently exploring  extensions beyond the standard model
   in which neutrinos couple to scalar fields motivated by Majoron models, in these extensions the coupling to the scalar (Majoron) provides
   a different   scale that permits to circumvent this  potential caveat. We expect to report on our results in a forthcoming article\cite{honew}.

 \section{Implications for sterile neutrino
 production in the early Universe:}\label{sec:implications}

\subsection{Time averaged transition probability, production rate and shortcomings of the rate
equation}\label{subsec:timeav}

The effective sterile production rate proposed in ref.\cite{foot}
and given by eqn. (\ref{effrate}) requires the average of the
transition probability $P_{a\rightarrow s}(t)$ over the decoherence
time scale. Hence, combining   (\ref{avedef}) with the transition
probability in the medium
(\ref{proboft},\ref{gammatotal},\ref{gammafin}) yields the following
time averaged transition
     probability (compare to eqn. (\ref{transaver}))

     \be \Big \langle P_{a\rightarrow s} \Big \rangle =    \frac{\sin^22\theta_m(k)}{2}~~ \frac{\cos^2 2\theta_m(k) +
     \Big(\frac{2\Delta E(k)}{  \Gamma_{aa}(k)}
     \Big)^2}{\sin^22\theta_m(k)
     \Bigg[1+ \Big(\frac{2\Delta E(k)}{  \Gamma_{aa}(k)}  \Big)^2 \Bigg]}
       \label{avefin} \ee

We have purposely kept the $\sin^22\theta_m(k)$ in the numerator
 and  denominator to highlight   the cancelation between this
 factor arising from the transition probability in the numerator
 with the factor $1-(\gamma/\Gamma)^2$ arising from the total
 integrated probability in the denominator. The factor  $\cos^22\theta_m(k)$ in the numerator and
the $\sin^22\theta_m(k)$  in the denominator are
 hallmarks of the presence of the \emph{two different relaxation rates} $\Gamma_1(k),\Gamma_2(k)$, and are responsible for the
 difference with the result (\ref{prob}). The extra factor $\sin^22\theta_m(k)$ in the denominator  signals an enhancement when
$\theta_m(k)=0,\pi/2$. In the  case $\theta_m(k)\sim 0$ the
relaxation rate $\Gamma_2(k) \ll \Gamma_1(k)$ whereas for
$\theta_m(k) \sim \pi/2$ the opposite holds, $\Gamma_1(k) \ll
\Gamma_2(k)$. In either case  there is a wide separation between the
relaxation rates of the propagating modes in the medium and the
longest time scale for relaxation   dominates the time integral in
(\ref{avefin}). This is depicted in fig. (\ref{fig:pas}).

This is an important difference with the result in \cite{foot}
wherein it was assumed that $\Gamma_1 =\Gamma_2$, in which case
$\gamma=0$. For $\theta_m(k) \sim 0,\pi/2$, the ratio
  $\gamma/\Gamma \sim 1$ leads to  an enhancement of the time averaged transition probability.
  The interpretation of this result should be clear. The probability
  $P_{a\rightarrow s}(t)$ has two distinct contributions,
  the interference oscillatory term, and the non-oscillatory
  terms. When one of these non-oscillatory terms features a much longer relaxation
  time scale, it dominates the integrand at long time  after the interference term has become
  negligible, as shown in
  figure (\ref{fig:pas}). Therefore the time integral receives the largest contribution from the term with the
  smallest relaxation rate, this is the origin of the factor $1-(\gamma/\Gamma)^2= \sin^22\theta_m(k)$ in the denominator.

Taking the kinetic equation that describes sterile neutrino
 production (\ref{kine}) along with the effective production rate (\ref{effrate}) \textbf{at face
 value},   the new result (\ref{avefin})  for the average
 transition probability yields the  effective production rate

 \be \label{ratef} \Gamma(a\rightarrow s;k) =
 \frac{\Gamma_{aa}(k)}{4}\, \frac{\cos^2 2\theta_m(k) +
     \Big(\frac{2\Delta E(k)}{   \Gamma_{aa}(k)}
     \Big)^2}{
     \Bigg[1+ \Big(\frac{2\Delta E(k)}{   \Gamma_{aa}(k)}  \Big)^2
     \Bigg]}\,. \ee

      The result of references\cite{foot,dibari,kev1} is retrieved
     \emph{only} near an MSW resonance for which $\cos2\theta_m(k)
     \approx 0$, in this case the relaxation rates become the same
     and $\gamma=0$. However, accounting for \emph{both} relaxation
     rates $\Gamma_1;\Gamma_2$ yields the new result (\ref{avefin},\ref{ratef})
     which is generally very different from the usual one
     (\ref{prob}).

 The result  (\ref{ratef}) is   in clear contradiction with the
 analysis in section (\ref{subsec:phys}) wherein the physical interpretation
 of the damping rates identify the sterile degrees of freedom as
 very weakly coupled to the plasma both at high temperature
   ($\theta_m \sim \pi/2$) and low temperature ($\theta_m \sim 0$),
   therefore should feature \emph{small} production rates. Contrary
   to this expectation,
 taking the limit of $\sin 2\theta_m(k) \sim 0$ in (\ref{ratef}),   still yields a non-vanishing sterile neutrino
 production rate despite the fact sterile neutrinos decouple from
 the plasma in this limit. The origin of this puzzling result is the {\bf
 time averaged probability} $\langle P_{a\rightarrow s}\rangle$
 (\ref{avedef}) and
 \emph{not    any ambiguity in the calculation of the relaxation rates or in the
 time dependence of the transition probability}
 $P_{a\rightarrow s}(t)$. The time integral in the averaged expression
 (\ref{avedef}) introduces a \emph{denominator} $\sin^22\theta_m(k)$
 from the longest time scale, and it is this denominator that is
 responsible for the enhancement. Thus the unreliability of the
 result  (\ref{ratef}) is a direct consequence of
 using the time-averaged transition probability (\ref{avedef}) in
 the rate equation (\ref{kine}).

 The real
 time  analysis presented above clearly suggests that far away from
 an MSW  resonance when $\Gamma_1$ and $\Gamma_2$ differ widely, $\Gamma^{-1}$ is \emph{not}
 the relevant  time scale  for suppression of the transition
 probability, but the \emph{longest} of $\Gamma^{-1}_1$ and
 $\Gamma^{-1}_2$ therefore the time averaged transition probability (\ref{avedef}) \emph{cannot
 be the correct ingredient in the rate equation}. A more suitable definition of the
 average transition probability under these circumstances should be
 \be \langle P_{a\rightarrow s} \rangle =
 \Gamma_{sm}\,\int^\infty_0 P_{a\rightarrow s}(t) dt
 \label{betterrate}\ee where $\Gamma_{sm}$ is the
 \emph{smallest of} $\Gamma_{1,2}$. In a non-expanding cosmology
 this would indeed be the correct definition of an average
 transition probability, however in the early Universe as the
 temperature diminishes upon cosmological expansion,
 $\Gamma_{sm}$ changes   with time   crossing from $\Gamma_1$ over to $\Gamma_2$ at the
 resonance and the alternative definition (\ref{betterrate}) would imply a ``rate'' with a sliding averaging time scale
 that changes rapidly near an MSW resonance.
  One can instead provide yet another suitable definition
 of an averaged transition rate \be   \langle P_{a\rightarrow s} \rangle =
  \frac{\Gamma_1\,\Gamma_2}{\Gamma_1+\Gamma_2}\,\int^\infty_0 P_{a\rightarrow s}(t)
  dt\,.
 \label{betterrate2}\ee When the two rates differ widely the prefactor always approximates the smaller one. Since \be
 \frac{\Gamma_1\,\Gamma_2}{\Gamma_1+\Gamma_2} =
 \frac{\Gamma_{aa}}{4}
 \left[1-\frac{\gamma^2(k)}{\Gamma^2(k)}\right] \ee this definition would cancel the  enhancement
 from the $\sin^22\theta_m(k)$ in the denominator in (\ref{avefin})
 (still leaving the $\cos^22\theta_m(k)$ in the numerator), but it
 misses the \emph{correct} definition of the average rate by a
 factor $2$, namely by $100\%$, in the region of the resonance where $\Gamma_1=\Gamma_2 =\Gamma_{aa}/2$.

 Obviously the ambiguity in properly defining a time averaged
 transition probability stems from the wide separation of the time
 scales associated with the damping of the quasiparticle modes,
 far away from an MSW resonance. Near the resonance both time scales become
 comparable and there is no ambiguity in the averaging scale. Complicating this issue further is the fact that in
 the early Universe these time scales are themselves time dependent
 as a consequence of the cosmological expansion and feature a rapid crossover behavior at an MSW resonance.

\vspace{2mm}

\subsection{Caveats of the kinetic description.}

It is important to highlight the main \emph{three different} aspects
at the origin of the enhanced production rate given by equation
(\ref{ratef}) in the high temperature regime, for $\theta_m(k) \sim
\pi/2$: i) the {\bf assumption} of the validity of the usual rate
equation in terms of a time-averaged transition probability wherein
the relevant time scale for averaging is the decoherence time scale
$1/\Gamma$, ii) the result of a complete self-energy calculation
that yields \emph{two time scales} which are widely different far
away from an MSW resonance, in particular at very high and very low
temperatures, iii) the generalized transition probability \emph{in
the medium} (\ref{proboft2}) obtained from expectation values in the
full density matrix, rather than the usual quantum mechanical
expression in terms of single particle states. The real time study
of the transition probability shows that the \emph{oscillatory
interference term}  is suppressed on the \emph{decoherence} time
scale $1/\Gamma$,  but also that this is \emph{not} the relevant
time scale for the suppression of the transition probability far
away from an MSW resonance. The transition probability actually
grows during $1/\Gamma$ reaches its maximum on this time scale and
remains near this value for a long time interval between $1/\Gamma$
and $1/\Gamma_{sm}$ where $\Gamma_{sm}$ is the \emph{smaller} of
$\Gamma_1,\Gamma_2$. The enhanced rate emerges when  \emph{taking
for granted} the definition of the time-averaged probability in
terms of the decoherence scale but including in this expression the
correct  form of the transition probability (\ref{proboft2}). As
discussed above, alternative definitions of a time-averaged rate
could be given, but all of them have caveats when applied to sterile
neutrino production in the early Universe.

However, we emphasize, that the underlying physical reason for the
enhancement does not call for a simple \emph{redefinition} of the
rate but for a full reassessment of the kinetic equation of sterile
neutrino production. The important fact is that the wide separation
of scales \emph{prevent} a consistent description in terms of a
simple \emph{rate} in the kinetic equation, a \emph{rate} implies
only one relevant time scale for the build-up or relaxation of
population, whereas our analysis reveals two widely different scales
that are of the same order only near an MSW resonance.

Kinetic rate equations are  generally a Markovian limit of   more
complicated equations in  which the  transition probability   in
general features a \emph{non-linear time dependence} . Only
 when the non-linear aspects of the time dependence of the transition probability  are transients that disappear
  faster than the scale of build-up or
 relaxation an average transition probability per unit time, namely a rate, can be defined and   the
 \emph{memory} aspects associated with the time evolution of the
 transition probability can be neglected. This     \emph{is not}   the case if there is
 a wide separation of scales, and under these circumstances  the assumptions
 leading to the kinetic equation (\ref{kine}) must be revised and
 its validity questioned, very likely requiring a reassessment of the kinetic description.
 This situation becomes even more pressing in the early Universe. In
 the derivation of the average probability in ref.\cite{foot} the
 rate $\Gamma_{aa}$ (denoted by $\tau_0$ in that reference) is
 taken as a constant in the time integral in the average. This is a
 suitable approximation \emph{if} the integrand falls off in the
 time  scale $1/\Gamma_{aa}$, since this time scale is shorter than
 the Hubble expansion time scale for $T > 1\,\textrm{MeV}$. However,
 if there is a \emph{much longer} time scale, when one of the
 relaxation rates is very small, as is the case depicted in fig.(\ref{fig:pas}), then this approximation cannot be
 justified and a full time-dependent kinetic description beyond a simple rate equation must be
 sought.

 Thus we are led to conclude that the simple rate equation
 (\ref{kine}) based on the time-averaged transition probability
 (\ref{avedef}) is likely \emph{incorrect} far away from MSW resonances.

 An alternative kinetic description based on a production rate
 obtained from   quantum field theory   has
 recently been offered\cite{shaposhnikov06} and seems to yield a result very different for the rate
  equation (\ref{kine}) in terms of the time-averaged transition probability. However,
   this alternative description   focuses
 on the hadronic contribution near the MSW resonance, and as such
 cannot yet address the issue of the widely separated time scales far away from it. A
 full quantum field theoretical treatment far away from an MSW
 resonance which systematically and consistently treats the two widely different time scales
 is not yet available.

 \vspace{2mm}

 Thus we conclude that while the result for the rate (\ref{ratef})
 is a \emph{direct consequence} of including the correct transition
 probability $P_{a\rightarrow s}(t)$ given by (\ref{proboft2}) into the
 rate equation (\ref{kine}),   our   field
 theoretical analysis of the full neutrino propagator in the medium,
 and the real time evolution of the transition probability,
 extracted from the \emph{full density matrix} leads us to challenge
 the validity of the simple  rate equation (\ref{kine}) with (\ref{effrate})
  to describe sterile neutrino production in the
 early Universe away from an MSW resonance.

\section{Conclusions:}\label{sec:conclu}

Motivated by the cosmological importance of sterile neutrinos,   we
reconsider an important aspect of the kinetics of sterile neutrino
production via active-sterile oscillations at high temperature:
quantum Zeno suppression of the sterile neutrino production rate.

Within an often used  kinetic approach to sterile neutrino
production, the   production rate involves two ingredients: the
active neutrino scattering rate $\Gamma_{aa}$ and a time  averaged
active-sterile transition probability $\langle P_{a\rightarrow s}
\rangle$\cite{kainu,cline,foot,dibari,kev1} in the case of one
sterile and one active neutrino.

Unlike the usual treatment in terms of a truncated $2\times 2$
density matrix for flavor degrees of freedom, we study the dynamics
of active-sterile transitions directly from the \emph{full real time
evolution of the quantum field density matrix}. Active-sterile
transitions are studied as an initial value problem wherein the main
ingredient is the \emph{full neutrino propagator in the medium},
obtained directly from the quantum density matrix and includes the
self-energy up to  $\mathcal{O}(G^2_F)$. The correct dispersion
relations and damping rates of the quasiparticles modes are obtained
from the neutrino propagator in the medium. We introduce a
generalization of the active-sterile transition probability from the
expectation values of the neutrino field operators in the density
matrix.

  There are three main results from our study:

  \begin{itemize}

  \item{ {\bf I):} The damping rates of the two different propagating modes in the medium are given by \be \Gamma_1(k)
=    \Gamma_{aa}(k)  \cos^2\theta_m(k)\,; \,\Gamma_2(k) =
\Gamma_{aa}(k)  \sin^2\theta_m(k) \ee where $\Gamma_{aa}(k) \propto
G^2_F \,k\, T^4$ is the active neutrino scattering rate and
$\theta_m(k)$ is the mixing angle in the medium. The dispersion
relations are the usual ones with the index of refraction
correction\cite{notzold}, plus perturbatively small two-loop
corrections of $\mathcal{O}(G^2_F)$. We give a simple physical
explanation for this result: for very high temperature when
$\theta_m \sim \pi/2$, $\nu_a \sim \nu_2;\nu_s \sim \nu_1$ and
$\Gamma_2 \sim \Gamma_{aa};\Gamma_1 \ll \Gamma_{aa}$. In the
opposite limit of very low temperature and small vacuum mixing angle
$\theta_m \sim 0$ , $\nu_a \sim \nu_1;\nu_s \sim \nu_2$ and
$\Gamma_1 \sim \Gamma_{aa};\Gamma_2 \ll \Gamma_{aa}$. Thus in either
case the sterile neutrino is much more weakly coupled to the plasma
than the active one. }

\item{{\bf II):} We study the active-sterile transition probability $P_{a\rightarrow s}(t)$ directly in real
time from the time evolution of expectation values of the neutrino
field operators in the density matrix.   The result is given by \be
P_{a\rightarrow s}(k;t) =
     \frac{\sin^22\theta_m(k)}{4} \,
     \Bigg[ {e^{-\Gamma_1(k)\,t}} + {e^{-\Gamma_2(k)\,t}} \,-  2\, e^{-\frac{1}{2}(\Gamma_1(k)+\Gamma_2(k))t}\cos(\Delta E(k) t)
     \Bigg] \,.
     \label{proboftcon} \ee  The
real time analysis shows that  even when  $\Gamma(k) \gg \Delta
E(k)$, which in the literature \cite{stodolsky,kev1} is taken to
indicate quantum Zeno suppression,    the transition probability is
substantial on time scales
     much longer than $\Gamma^{-1}$ if $\Gamma_1$ and $\Gamma_2$ are \emph{widely
     separated}. While the oscillatory interference term is suppressed by the \emph{decoherence} time scale
     $1/\Gamma(k)=2/\Gamma_{aa}(k)$ in agreement with the results of
     \cite{stodolsky,foot},  at very high or low temperature this
     is \emph{not} the relevant time scale for the suppression of
     the transition probability, which is given by $1/\Gamma_{sm}$
     with $\Gamma_{sm}$ the \emph{smaller} between
     $\Gamma_1,\Gamma_2$. We obtain the \emph{complete} conditions for quantum Zeno suppression: i) $2\Delta E(k)/\Gamma_{aa} \ll 1$ where
     $\Delta E(k)$ is the oscillation frequency in the medium, \emph{and} ii) $\Gamma_1 \sim \Gamma_2$. This latter condition is only achieved
     near an MSW resonance. Furthermore we studied consistently up to second order in standard model weak
interactions,  in which temperature regime the quantum Zeno
condition $\Gamma_{aa}(k) \gg \Delta E(k)$ is fulfilled. We find
that for $m_s \sim \textrm{keV}$ and $10^{-5} \lesssim \sin2\theta
\lesssim 10^{-3}$\cite{dodelson,kev1,dibari} the \emph{opposite}
condition, $\Gamma_{aa}(k) \ll \Delta E(k)$ is fulfilled in the high
temperature limit $T\gg T_{MSW}\sim 215 \,\textrm{MeV}$, as well as
in the \emph{low} temperature regime $3\,\textrm{MeV}\lesssim T \ll
T_{MSW}$. We therefore conclude that the quantum Zeno conditions are
\emph{may only} be fulfilled near an MSW resonance for $T\sim
T_{MSW}$ but a firmer assessment of this possibility requires to
include the $\mathcal{O}(G^2_F)$ corrections to the index of
refraction. }

\item{ {\bf III):} Inserting the result (\ref{proboftcon}) into the expressions for the time averaged transition probability (\ref{avedef}) and
the sterile neutrino production rate (\ref{effrate}) yields an
expression for this rate that is \emph{enhanced} at very high or low
temperature given by equation (\ref{ratef}) instead of the result
(\ref{prob}) often used in the literature. The surprising
enhancement at high or low temperature implied by (\ref{ratef})
originates in \emph{two} distinct aspects: i) the \emph{assumption}
of the validity of the rate kinetic equation in terms of a
time-averaged transition probability with an
 averaging time scale determined by the decoherence scale $2/\Gamma_{aa}$, and ii) inserting the result (\ref{proboftcon}) into the
 definition of the time-averaged transition probability. The enhancement is a distinct result of the \emph{fact} that at
 very high or low temperatures the decoherence time scale is \emph{not} the relevant scale for suppression of $P_{a\rightarrow s}$ but either
 $1/\Gamma_1$ or $1/\Gamma_2$ whichever is \emph{longer} in the appropriate temperature regime. Our analysis shows that far away from the
 region of MSW resonance, the transition probability reaches its maximum on  time scale $1/\Gamma(k)$,
  remains  near this value  during  a long time   scale $1/\Gamma_{sm} \gg 1/\Gamma$.
  We have also argued that in the early Universe the definition
 of a time averaged transition probability is \emph{ambiguous} far away from MSW resonances.
Our analysis leads us to conclude that the simple rate equation
(\ref{kine}) in terms of the production rate (\ref{effrate}), (
\ref{avedef}) is likely incorrect far away from MSW resonances. }

  \end{itemize}

 We emphasize and clarify an important distinction between the results summarized above.
 Whereas  ${\bf I}$ and ${\bf II}$ are solidly based on a consistent and systematic quantum
 field theory calculation of the neutrino propagator, the correct
 equations of motion for the quasiparticle modes in the medium and the time evolution of
 expectation values of neutrino field operators in the  quantum density matrix, the
 results summarized in ${\bf III}$ are a consequence of the
 {\bf assumption} on the validity of the kinetic description based
 on the simple rate equation (\ref{kine}) with an effective rate
 (\ref{effrate}) in
 terms of the time-averaged transition probability (\ref{avedef}).
 The enhancement of the sterile production rate arising from this
 {\bf assumption}, along with the ambiguity in properly defining a
 time-averaged transition probability in an expanding cosmology in the temperature
 regime far away from a MSW resonance all
 but suggest  important caveats in the
validity of  the kinetic description for sterile neutrino production
in terms of a simple rate equation in this regime.   Our analysis
suggests that a deeper understanding of possible quantum Zeno
suppression at high temperature requires a reassessment of the
validity of the perturbative expansion in the standard model or in
Fermi's effective field theory. Further studies of these issues are
in progress.

 \acknowledgments The authors thank Kev Abazajian and Scott Dodelson for enlightening discussions,
  comments and suggestions, Micha Shaposhnikov for correspondence and Georg Raffelt for correspondence and probing
  questions. They acknowledge support
 from the National Science Foundation through grant awards:
 PHY-0242134,0553418. C. M. Ho acknowledges partial support through
 the Andrew Mellon Foundation and the Daniels Fellowship.


\begin{thebibliography}{99}
\bibitem{book1} C. W. Kim and A. Pevsner, \textit{Neutrinos in Physics and
Astrophysics}, (Harwood Academic Publishers, USA, 1993).

\bibitem{book2} R. N. Mohapatra and P. B. Pal, \textit{Massive Neutrinos in
Physics and Astrophysics}, (World Scientific, Singapore, 2004).

\bibitem{book3} M. Fukugita and T. Yanagida, \textit{Physics of Neutrinos
and Applications to Astrophysics}, (Springer-Verlag Berlin
Heidelberg 2003).

\bibitem{raffeltbook} G. G. Raffelt, \textit{Stars as Laboratories for
Fundamental Physics}, (The University of Chicago Press, Chicago,
1996); astro-ph/0302589; New Astron.Rev. \textit{46}, 699 (2002);
 hep-ph/0208024.

\bibitem{dodelson} S. Dodelson and L. M. Widrow, Phys. Rev. Lett.
\textbf{72}, 17 (1994).

\bibitem{asaka} T. Asaka, M. Shaposhnikov, A. Kusenko, Phys. Lett.
\textbf{B 638}, 401 (2006).

\bibitem{shi} X. Shi, G. M. Fuller, Phys. Rev. Lett. \textbf{83},
3120 (1999).

\bibitem{kev1} K. Abazajian, G. M. Fuller, M. Patel, Phys. Rev.
\textbf{D64}, 023501 (2001).

\bibitem{hansen} A. D. Dolgov and S.  H. Hansen, Astropart. Phys.
\textbf{16}, 339 (2002).

\bibitem{kev2} K. Abazajian, G. M. Fuller, Phys. Rev. \textbf{D66},
023526 (2002).

\bibitem{kev3} K. Abazajian, Phys. Rev. \textbf{D73}, 063506 (2006), \emph{ibid}, 063513 (2006).

\bibitem{kusenko} P. Biermann, A. Kusenko, Phys. Rev. Lett.
\textbf{96}, 091301 (2006).

\bibitem{kou} K. Abazajian, S. M. Koushiappas, Phys. Rev.
\textbf{D74} 023527 (2006).

\bibitem{dolgovrev} A. D. Dolgov,
Phys. Rept. \textbf{370}, 333 (2002);  Surveys High Energ.Phys.
\textbf{17} 91 (2002).

\bibitem{pastor} J. Lesgourgues, S. Pastor, Phys.Rept. \textbf{429 }
307, (2006).

\bibitem{raffeltSN} G.~Raffelt, G.~Sigl, Astropart.Phys. \textbf{1}, 165
(1993).

\bibitem{hannestad} S. Hannestad, hep-ph/0602058.

\bibitem{bier} P. L. Biermann, F.l Munyaneza,astro-ph/0609388;
astro-ph/0702164;astro-ph/0702173;

\bibitem{fuller} J. Hidaka, G. M. Fuller, astro-ph/0609425.

\bibitem{fuller2} C. J. Smith, G. M. Fuller, C. T. Kishimoto, K.
Abazajian, astro-ph/0608377.

\bibitem{fuller3} C. T. Kishimoto, G. M. Fuller, C. J. Smith,
astro-ph/0607403.

\bibitem{segre} A. Kusenko, G. Segre, Phys. Rev. \textbf{D59},
061302, (1999).

\bibitem{fullkus} G. M. Fuller, A. Kusenko, I. Mociouiu, S. Pascoli,
Phys. Rev. \textbf{D68}, 103002 (2003).


 \bibitem{miniboone} A. A. Aguilar-Arevalo \textit{et.al.}
 (MiniBooNE collaboration) arXiv:0704.1500 [hep-ex].

 \bibitem{lsnd1} C. Athanassopoulos \emph{et.al.} (LSND
collaboration), Phys.Rev.Lett. \textbf{81}, 1774 (1998).

\bibitem{lsnd2} A. Aguilar \emph{et.al.} (LSND
collaboration), Phys.Rev. \textbf{D64 },  112007 (2001).

\bibitem{malto} M. Maltoni, T. Schwetz, arXiv:0705.0107 [hep-ph].

\bibitem{Xray} K. Abazajian, G. M. Fuller, W. H. Tucker, Astrop. J.
\textbf{562}, 593 (2001).

\bibitem{boyarsky} A. Boyarsky, A Neronov, O Ruchayskiy, M.
Shaposhnikov, Mon.Not.Roy.Astron.Soc. \textbf{370}, 213 (2006); JETP
Lett.\textbf{ 83 }, 133 (2006); Phys.Rev. \textbf{D74}, 103506
(2006); A. Boyarsky, A. Neronov, O. Ruchayskiy, M. Shaposhnikov, I.
Tkachev, astro-ph/0603660; A. Boyarsky, J. Nevalainen, O.
Ruchayskiy, astro-ph/0610961; A. Boyarsky, O. Ruchayskiy, M.
Markevitch, astro-ph/0611168.

\bibitem{hansen2} S.~Riemer-Sorensen, K.~Pedersen, S.~H.~Hansen, H.~Dahle,
 arXiv:astro-ph/0610034;
 S.~Riemer-Sorensen, S.~H.~Hansen and K.~Pedersen,
 Astrophys.\ J.\  {\bf 644} (2006) L33
 (arXiv:astro-ph/0603661).

\bibitem{kou2} K. N. Abazajian, M. Markevitch, S. M. Koushiappas, R. C.
Hickox, astro-ph/0611144.

\bibitem{shapolast} F. Bezrukov, M. Shaposhnikov, hep-ph/0611352.

\bibitem{misra} B. Misra, E. C. C. Sudarshan, J. Math. Phys.
\textbf{18}, 756 (1977).

\bibitem{stodolsky} R. A. Harris, L. Stodolsky, Phys. Lett.
\textbf{B116}, 464 (1982); L. Stodolsky, Phys.Rev.
\textbf{D36}:2273,1987.

\bibitem{sigl} G. Raffelt, G. Sigl, L. Stodolsky, Phys. Rev. Lett.
\textbf{70}, 2363 (1993).

\bibitem{raffkin} G.Sigl and G.Raffelt,   Nucl.Phys.\textbf{B
406}, 423 (1993).

\bibitem{dolgov} A. Dolgov, Sov. J. Nucl. Phys. \textbf{33}, 700
(1981); R. Barbieri, A. Dolgov, Nucl. Phys. \textbf{B349}, 743
(1991),  (see also \cite{dolgovrev}).

\bibitem{enquist} K. Enqvist, K. Kainulainen, J. Maalampi, Nucl.
Phys. \textbf{B349}, 754 (1991); Phys. Lett.  \textbf{B244}, 186
(1990); K. Enqvist, K. Kainulainen, M. Thompson, Nucl. Phys.
\textbf{B373}, 498 (1992).


\bibitem{kainu} K. Kainulainen, Phys. Lett. \textbf{B244}, 191
(1990).

\bibitem{cline} J. Cline, Phys. Rev. Lett. \textbf{68}, 3137 (1992).

\bibitem{foot} R. Foot, R. R. Volkas, Phys. Rev. \textbf{D55}, 5147
(1997).

\bibitem{dibari} P. Di Bari, P. Lipari, M. Lusignoli, Int. J. Mod.
Phys. \textbf{A15}, 2289 (2000).

\bibitem{volkas} R. R. Volkas, Y. Y. Y. Wong, Phys. Rev.
\textbf{D62}, 093024 (2000); K. S. M. Lee, R. R. Volkas, Y. Y. Y.
Wong, \emph{ibid} 093025 (2000).

\bibitem{shaposhnikov06} T. Asaka, M. Laine and  M. Shaposhnikov, JHEP 0606 (2006)
053; JHEP 0701 (2007) 091.

\bibitem{raffelt} S. Blanchet, P. Di Bari, G.G. Raffelt,
hep-ph/0611337.

\bibitem{recentours}   D. Boyanovsky, C. M. Ho, Phys. Rev. \textbf{D75}, 085004 (2007).

\bibitem{notzold} D. Notzold and G. Raffelt, Nucl. Phys.
\textbf{B307}, 924 (1988).

\bibitem{charged} D. Boyanovsky, C. M. Ho, Astropart.Phys. 27 (2007)
99.

\bibitem{schwinger} J. Schwinger, J. Math. Phys. \textbf{2}, 407
(1961).

\bibitem{maha} K. T. Mahanthappa, Phys. Rev. \textbf{126}, 329
(1962); P. M. Bakshi and K. T. Mahanthappa, J. Math. Phys.
\textbf{41}, 12 (1963).

\bibitem{keldysh} L. V. Keldysh, JETP \textbf{20}, 1018 (1965).

\bibitem{revi} See also the following reviews:  K. Chou, Z. su, B. Hao and L. Yu, Phys. Rept.
\textbf{118}, 1 (1985); A. Niemi and G. Semenoff, Ann. Phys. (NY),
\textbf{152}, 105 (1984); N. P. Landsmann and C. G. van Weert, Phys.
Rept. \textbf{145}, 141 (1985); J. Rammer and H. Smith, Rev. of Mod.
Phys. \textbf{58}, 323 (1986).

\bibitem{eunu} C. M. Ho, D. Boyanovsky, H. J. de Vega, Phys.Rev. \textbf{D72
}, 085016 (2005); C. M. Ho, D. Boyanovsky, Phys.Rev. \textbf{D73},
125014 (2006).






\bibitem{fetter} A. Fetter and D. Walecka, \textit{Quantum Theory of
Many Particle Systems}, (McGraw-Hill, San Francisco 1971); G. D.
Mahan, \textit{Many Particle Physics}, (Plenum Press, New York,
1990).

\bibitem{ftf} J. I. Kapusta, \textit{Finite Temperature Field
Theory}, (Cambridge Monographs on Mathematical Physics, Cambridge
University Press 1989); M. Le Bellac, \textit{Thermal Field Theory}
 (Cambridge Monographs on Mathematical Physics, Cambridge University
Press 1996).



\bibitem{disip}D. Boyanovsky, H. J. de Vega and R. Holman, Proceedings of
the Second Paris Cosmology Colloquium, Observatoire de Paris, June
1994, pp.~127-215, H. J. de Vega and N. Sanchez, Editors (World-
Scientific, 1995); Advances in Astrofundamental Physics, Erice
Chalonge Course, N. Sanchez and A. Zichichi Editors, (World
Scientific, 1995); D. Boyanovsky, H. J. de Vega, R. Holman, D.-S.
Lee and A. Singh, Phys. Rev. {\bf D51}, 4419 (1995); D. Boyanovsky,
H. J. de Vega, R. Holman and J. Salgado, Phys. Rev. {\bf D54}, 7570
(1996). D. Boyanovsky, H. J. de Vega, C. Destri, R. Holman and J.
Salgado, Phys. Rev. {\bf D57}, 7388 (1998).

\bibitem{tadpole} D. Boyanovsky, H. J. de Vega and R. Holman, Proceedings of
the Second Paris Cosmology Colloguium, Observatoire de Paris, June
1994, pp.~127-215, H. J. de Vega and N. Sanchez, Editors (World
Scientific, 1995); Advances in Astrofundamental Physics, Erice
Chalonge Course, N. Sanchez and A. Zichichi Editors, (World
Scientific, 1995); D. Boyanovsky, H. J. de Vega, R. Holman and D.-S.
Lee, Phys. Rev. {\bf D52}, 6805 (1995).

\bibitem{nosfermions} S. Y.-Wang, D. Boyanovsky, H. J. de Vega,
D.-S. Lee and Y. J. Ng, Phys. Rev. \textbf{D61}, 065004 (2000); D.
Boyanovsky, H. J. de Vega, D.-S.Lee, Y.J. Ng and S.-Y. Wang, Phys.
Rev. \textbf{D59}, 105001 (1999).

\bibitem{weldon} H. A. Weldon, Phys. Rev. \textbf{D26}, 2789 (1982);
Phys. Rev. \textbf{D28}, 2007 (1983); Phys. Rev. \textbf{D40}, 2410
(1989).

\bibitem{kaons} See the recent review by R. Fleischer, hep-ph/0608010.

\bibitem{raiden} For a thorough pedagogical description see: A. Seiden, \textit{Particle Physics: A comprehensive
Introduction}, Addison Wesley, (San Francisco, 2004).




\bibitem{honew} D. Boyanovky, C. M. Ho, in preparation.




\end{thebibliography}
\end{document}